\RequirePackage{silence}
\PassOptionsToPackage{table}{xcolor}
\documentclass[preprint]{vldb}
\usepackage[techreport,vldb,libertine,savetrees,camera,arxiv]{manuprep}
\ifvldb
\fi

\renewcommand{\baselinestretch}{0.95}

\ifvldb
\fi

\usepackage{subcaption}
\usepackage[font=small, labelfont=bf]{caption}
\usepackage{float}
\usetikzlibrary{arrows,decorations.pathreplacing,calc,shapes,positioning,tikzmark}

\input{algofix.tex}

\xspaceaddexceptions{[]\}()}

\allowdisplaybreaks

\pgfplotsset{
  width=6cm,
  height=4cm, 
  compat=newest,
  xlabel near ticks,
  ylabel near ticks
}

\makeatletter
\newcommand{\gettikzxy}[3]{%
  \tikz@scan@one@point\pgfutil@firstofone#1\relax
  \edef#2{\the\pgf@x}%
  \edef#3{\the\pgf@y}%
}
\makeatother

\newcommand{\connectzm}[2]{
\draw[#2]
  ([shift={(-1.5pt,1.3ex)}]pic cs:#1-begin) 
    rectangle 
  ([shift={(0.5pt,-0.5ex)}]pic cs:#1-end);%
}

\ifthesis
\else
\definecolor{unsafe}{RGB}{254,0,82}
\definecolor{refresh}{RGB}{0,197,158}
\fi
\mdfsetup{usetwoside=false}

\lstdefinestyle{jupyterstyle}{%
  language         = python,
  basicstyle       = \scriptsize\ttfamily\bfseries,
  escapechar       = !,
  upquote          = true,
  tabsize          = 1,     
  numbers          = none,  
  commentstyle     = \color{pinegreen},      
  keywordstyle     = \color{blue},       
  stringstyle      = \color{red},    
  backgroundcolor=\transparent{0}\color{white},
  showstringspaces = false,              
  literate=%
    {0}{{{\color{magenta}{0}}}}1
    {1}{{{\color{magenta}{1}}}}1
    {2}{{{\color{magenta}{2}}}}1
    {3}{{{\color{magenta}{3}}}}1
    {4}{{{\color{magenta}{4}}}}1
    {5}{{{\color{magenta}{5}}}}1
    {6}{{{\color{magenta}{6}}}}1
    {7}{{{\color{magenta}{7}}}}1
    {8}{{{\color{magenta}{8}}}}1
    {9}{{{\color{magenta}{9}}}}1
    {True}{{{\color{pinegreen}{True}}}}1
    {False}{{{\color{pinegreen}{False}}}}1
}

\makeatletter
\lst@Key{celltype}{input}{\def\ATOcelltype{#1}}
\lst@Key{highlight}{white}{\def\ATOhighlight{#1}}
\lst@Key{cellnum}{}{\def\ATOcellnum{#1}}
\lst@Key{cellnumoffs}{\thesissub{0pt}{-65pt}}{\def\ATOcellnumoffs{#1}}
\lst@Key{cellnumfill}{red!20}{\def\ATOcellnumfill{#1}}
\lst@Key{leftmargin}{\thesissub{7pt}{72pt}}{\def\ATOleftmargin{#1}}
\lst@Key{rightmargin}{\thesissub{0pt}{65pt}}{\def\ATOrightmargin{#1}}
\makeatother

\lstnewenvironment{lstjupyter}[2][]{%
\lstset{#1}
\ifvldb
\vspace{-1em}
\fi
\tikzmark{sidebartop-#2}%
\mdframed[%
  roundcorner=3pt,
  leftmargin=\ATOleftmargin,
  innerleftmargin=0pt,
  innerrightmargin=0pt,
  rightmargin=\ATOrightmargin,
  innertopmargin=0pt,
  innerbottommargin=0pt,
  backgroundcolor=none
]%
\lstset{style=jupyterstyle,name=#2}
\newcounter{rectmark-#2}
\setcounter{rectmark-#2}{0}
\newcounter{rectangle-#2}
\setcounter{rectangle-#2}{0}
\def\HB{\stepcounter{rectangle-#2}\tikzmark{#2-\csname therectangle-#2\endcsname-begin}}
\def\HE{\tikzmark{#2-\csname therectangle-#2\endcsname-end}}
\def\HL{\stepcounter{rectmark-#2}\tikzmark{#2-\csname therectmark-#2\endcsname-rectmark}}
}{%
\endmdframed
\tikzmark{sidebarbot-#2}
\lstset{#1}
\ifnum\pdfstrcmp{\ATOcelltype}{output}=0

\else

\fi
\colorlet{highlight}{\ATOhighlight}
\begin{tikzpicture}[remember picture, overlay,
    every node/.append style={%
    	align=center,
    	minimum height=10pt,
        font=\scriptsize\ttfamily\bfseries,
    }%
]
\gettikzxy{(pic cs:line-#2-1-end)}{\ax}{\ay}
\ifvldb
\node[anchor=east,fill=\ATOcellnumfill] (#2-cell-marker) at ($(\linewidth, \ay) + (-6mm,0.5mm) + (\ATOcellnumoffs,0mm)$) {\ATOcellnum};
\else
\ifthesis
\node[anchor=east,fill=\ATOcellnumfill] (#2-cell-marker) at ($(\linewidth, \ay) + (-10mm,0.5mm) + (\ATOcellnumoffs,0mm)$) {\ATOcellnum};
\else
\node[anchor=east,fill=\ATOcellnumfill] (#2-cell-marker) at ($(\linewidth, \ay) + (-45mm,0.5mm) + (\ATOcellnumoffs,0mm)$) {\ATOcellnum};
\fi
\fi
\iftech 
\ifthesis
\fill[highlight,rounded corners] ([shift={(23mm,-3mm)}]pic cs:sidebartop-#2) rectangle ([shift={(21mm,8.5mm)}]pic cs:sidebarbot-#2);
\else
\fill[highlight,rounded corners] ([shift={(2mm,-2mm)}]pic cs:sidebartop-#2) rectangle ([shift={(0mm,6mm)}]pic cs:sidebarbot-#2);
\fi
\fi
\ifnum\csname therectmark-#2\endcsname>2
\foreach \k in {1,3,...,\csname therectmark-#2\endcsname}{%
	\draw[purple]
	  ([shift={(-1.5pt,1.3ex)}]pic cs:#2-\k-rectmark) 
	    rectangle 
	  ([shift={(0.5pt,-0.5ex)}]pic cs:#2-\number\numexpr\k+1\relax-rectmark);%
}%
\else
  \ifnum\csname therectmark-#2\endcsname>0
	\draw[purple]
	  ([shift={(-1.5pt,1.3ex)}]pic cs:#2-1-rectmark) 
	    rectangle 
	  ([shift={(0.5pt,-0.5ex)}]pic cs:#2-2-rectmark);%
  \fi
\fi
\end{tikzpicture}
\ifvldb
\vspace{-2.75em}
\else
\ifthesis
\vspace{-3.5em}
\else
\vspace{-2.0em}
\fi
\fi
}

\def\makesecondcellmarker#1#2{\begin{tikzpicture}[remember picture, overlay,
    every node/.append style={%
      align=center,
      minimum height=10pt,
        font=\scriptsize\ttfamily\bfseries,
        fill=blue!20
    }%
]
\gettikzxy{(pic cs:line-#1-1-end)}{\ax}{\ay}
\ifvldb
\node[anchor=east] (#1-second-marker) at ($(\linewidth, \ay) + (-6mm,0.5mm)$) {[#2]};
\else
\ifthesis
\node[anchor=east] (#1-second-marker) at ($(\linewidth, \ay) + (-31mm,0.5mm)$) {[#2]};
\else
\node[anchor=east] (#1-second-marker) at ($(\linewidth, \ay) + (-45mm,0.5mm)$) {[#2]};
\fi
\fi
\end{tikzpicture}}

\iftech
\makeatletter
\def\@copyrightspace{\relax}
\def\@copyrightblurb{\relax}
\def\@mkbibcitation{\relax}
\makeatother
\fi

\newcommand{\na}{\textsf{N/A}}
\makeatletter
\newcommand*{\phd}{%
    \@ifnextchar{.}%
        {Ph.D}%
        {Ph.D.\@\xspace}%
}
\makeatother

\makeatletter
\newcommand*{\prof}{%
    \@ifnextchar{.}%
        {Prof}%
        {Prof.\@\xspace}%
}
\makeatother

\makeatletter
\newcommand*{\dr}{%
    \@ifnextchar{.}%
        {Dr}%
        {Dr.\@\xspace}%
}
\makeatother

\makeatletter
\newcommand*{\etc}{%
    \@ifnextchar{.}%
        {etc}%
        {etc.\@\xspace}%
}
\makeatother

\makeatletter
\newcommand*{\etal}{%
    \@ifnextchar{.}%
        {et al}%
        {et al.\@\xspace}%
}
\makeatother

\makeatletter
\newcommand*{\resp}{%
    \@ifnextchar{.}%
        {resp}%
        {resp.\@\xspace}%
}
\makeatother

\makeatletter
\newcommand*{\whp}{%
    \@ifnextchar{.}%
        {w.h.p}%
        {w.h.p.\@\xspace}%
}
\makeatother

\makeatletter
\newcommand*{\iid}{%
    \@ifnextchar{.}%
        {i.i.d}%
        {i.i.d.\@\xspace}%
}
\makeatother

\makeatletter
\newcommand*{\lhs}{%
    \@ifnextchar{.}%
        {L.H.S}%
        {L.H.S.\@\xspace}%
}
\makeatother

\makeatletter
\newcommand*{\rhs}{%
    \@ifnextchar{.}%
        {R.H.S}%
        {R.H.S.\@\xspace}%
}
\makeatother

\makeatletter
\newcommand*{\wrt}{%
    \@ifnextchar{.}%
        {w.r.t}%
        {w.r.t.\@\xspace}%
}
\makeatother

\makeatletter
\newcommand*{\sthat}{%
    \@ifnextchar{.}%
        {s.t}%
        {s.t.\@\xspace}%
}
\makeatother

\newcommand*{\eg}{e.g.\@\xspace}
\newcommand*{\ie}{i.e.\@\xspace}

\newcommand*{\naive}{na\"ive\xspace}
\newcommand*{\naively}{na\"ively\xspace}

\newenvironment{densecenter}{%
  \setlength\topsep{0pt}
  \setlength\parskip{0pt}
  \begin{center}
}{%
  \end{center}
}

\mdfdefinestyle{MyFrame}{%
  linecolor=gray,
  outerlinewidth=0.25pt,
  roundcorner=3pt,
  innertopmargin=\baselineskip,
  innerbottommargin=\baselineskip,
  innerrightmargin=10pt,
  innerleftmargin=10pt,
  backgroundcolor=white
}

\newenvironment{denselist}{
    \begin{list}{\small{$\bullet$}}%
    {\setlength{\itemsep}{0ex} \setlength{\topsep}{0ex}
    \setlength{\parsep}{0pt} \setlength{\itemindent}{0pt}
    \setlength{\leftmargin}{1.5em}
    \setlength{\partopsep}{0pt}}}%
    {\end{list}}

\newenvironment{denseenum}{
  \begin{enumerate}[
  wide=0pt,
  widest=99,
  leftmargin=\parindent,
  labelsep=*,
  topsep=2pt,
  partopsep=0pt,
  parsep=0pt,
  itemsep=1pt,
  ]
}{\end{enumerate}}

\newcommand{\topic}[1]{\vspace{-3.5pt}\smallskip \smallskip \noindent{\bf #1.}}

\newcommand{\emtitle}[1]{\vspace{0.3em}\noindent{\em #1}}
\newcommand{\frameme}[1]{
\thesissub{\noindent\fbox{
  \parbox{0.95\linewidth}{
    \noindent #1
    }
  }}{\begin{mdframed}#1\end{mdframed}}
}

\definecolor{pinegreen}{rgb}{0.0, 0.47, 0.44}
\definecolor{persiangreen}{rgb}{0.0, 0.65, 0.58}
\definecolor{officegreen}{rgb}{0.0, 0.5, 0.0}

\newcolumntype{C}[1]{>{\centering\arraybackslash}m{#1}}
\newcolumntype{M}[1]{>{\arraybackslash}m{#1}}
\newcolumntype{$}{>{\global\let\currentrowstyle\relax}}
\newcolumntype{^}{>{\currentrowstyle}}
\newcolumntype{!}{<{\gdef\nonheader{1}}}

\makeatletter
\newcolumntype{S}[1]{>{\arraybackslash\@ifundefined{nonheader}{}{#1}}c}
\makeatother
\newcommand{\undefnonheader}{\global\let\nonheader\relax}

\NewEnviron{curlymatrix}{%
  \LEFTRIGHT\lcbrace\rcbrace{\,\begin{matrix}\BODY\end{matrix}\,}%
}
\makeatletter
\NewEnviron{curlycases}{%
  \LEFTRIGHT\lcbrace.{%
    \,\array{@{}l@{\quad}l@{}}\BODY\endarray\,
  }%
}
\makeatother

\newenvironment{algo}{%
  \algorithm
}{%
  \endalgorithm
}

\makeatletter
\newcommand{\nosemic}{\renewcommand{\@endalgocfline}{\relax}}
\newcommand{\dosemic}{\renewcommand{\@endalgocfline}{\algocf@endline}}
\let\oldnl\nl
\newcommand{\nonl}{\renewcommand{\nl}{\let\nl\oldnl}}
\makeatother

\newcommand{\backref}{\rotatebox[origin=c]{90}{$\Rsh$}}

\newcommand{\BIT}{\begin{itemize}}
\newcommand{\EIT}{\end{itemize}}
\newcommand{\BNUM}{\begin{enumerate}}
\newcommand{\ENUM}{\end{enumerate}}
\newcommand{\bcircle}[1]{\ding{\number\numexpr 201 + #1 \relax}}
\newcommand{\cmark}{\ding{51}}%
\newcommand{\xmark}{\ding{55}}%
\newcommand{\rxmark}{\red{\xmark}}%
\newcommand\mbb[1]{\mathbb{#1}}

\def\bigo#1{\mathcal{O}\left(#1\right)} 
\def\indic#1{\mbb{I}\left\{{#1}\right\}} 

\def\E{\mathbb{E}} 
\def\Earg#1{\E\left[{#1}\right]}

\def\P{\mathbb{P}} 
\def\Parg#1{\P\left({#1}\right)}

\declaretheoremstyle[
notefont=\normalfont, notebraces={[}{]},
bodyfont=\normalfont\itshape,
headformat=\NAME~\NUMBER\NOTE
]{brackets}
\ifthesis
\declaretheorem[numberwithin=chapter,style=brackets]{definition}
\declaretheorem[numberwithin=chapter]{theorem}
\declaretheorem[numberwithin=chapter]{lemma}
\declaretheorem[numberwithin=chapter]{hypothesis}
\declaretheorem[numberwithin=chapter]{corollary}
\declaretheorem[numberwithin=chapter]{example}
\declaretheorem[numberwithin=chapter]{guarantee}
\declaretheorem[numberwithin=chapter]{constraint}
\declaretheorem[numberwithin=chapter]{problem}
\else
\declaretheorem[style=brackets]{definition}
\newtheorem{theorem}{Theorem}

\fi

\crefname{guarantee}{Guarantee}{Guarantees}
\crefname{problem}{Problem}{Problems}
\crefname{example}{Example}{Examples}
\crefname{constraint}{Constraint}{Constraints}

\newcommand\AVG{\textsf{AVG}\xspace}

\newcommand\code[1]{{\texttt{{\upshape#1}}}}

\newcommand\codespace[1]{\texttt{{\upshape#1}}\,\,}
\newcommand\spacecodespace[1]{\,\,\texttt{{\upshape#1}}\,\,}

\newcommand{\pp}{\mathcal{P}\xspace}

\newcommand{\nbs}{{\sc nbsafety}\xspace}

\renewcommand{\emptyset}{\varnothing}

\def\name{\textsc{nbsafety}\xspace}

\def\numrepos{$712$\xspace}
\def\numhistories{$657$\xspace}
\def\numsessions{$51000$\xspace}
\def\numrepaired{$2566$\xspace}
\def\numworking{$666$\xspace}
\def\numsafe{$549$\xspace}
\def\numunsafe{$117$\xspace}

\def\excthresh{$50$\%\xspace}
\def\leventhresh{$80$\%\xspace}

\ifthesis
\else
\def\unsafehl#1{{\bf\color{unsafe}\ul{#1}}}

\def\refresherhl#1{{\bf\color{refresh}\ul{#1}}}
\fi

\def\hlstale#1{#1}
\def\hlfresh#1{#1}
\def\hlrefresher#1{#1}

\def\metric{predictive power\xspace}
\def\hl{\mathcal{H}}
\def\pp{\mathcal{P}}
\def\univ{\mathcal{U}}

\def\hlf{\hlfresh{\mathcal{H}_f}}
\def\hlr{\hlrefresher{\mathcal{H}_{r}}}
\def\hls{\hlstale{\mathcal{H}_{s}}}
\def\hln{\mathcal{H}_n}

\def\hlrr{\mathcal{H}_{\text{\upshape{rnd}}}}

\def\hlstar{\mathcal{H}_*}

\def\hlfp{\hlfresh{\Delta\mathcal{H}_f}}
\def\hlrp{\hlrefresher{\Delta\mathcal{H}_{r}}}
\def\hlsp{\hlstale{\Delta\mathcal{H}_{s}}}
\newcommand{\hlfpt}[1][t]{\hlfresh{\Delta\mathcal{H}^{(#1)}_f}}
\newcommand{\hlrpt}[1][t]{\hlrefresher{\Delta\mathcal{H}^{(#1)}_{r}}}

\newcommand{\hlft}[1][t]{\hlfresh{\mathcal{H}^{(#1)}_f}}
\newcommand{\hlrt}[1][t]{\hlrefresher{\mathcal{H}^{(#1)}_{r}}}
\newcommand{\hlst}[1][t]{\hlstale{\mathcal{H}^{(#1)}_{s}}}

\def\live{\textsf{\upshape{\scriptsize{LIVE}}}\xspace}
\def\dead{\textsf{\upshape{\scriptsize{DEAD}}}\xspace}
\def\inited{\textsf{\upshape{\scriptsize{INITIALIZED}}}\xspace}
\def\invdead{\textsf{\upshape{\scriptsize{DEAD}}}^{-1}\xspace}
\def\use{\textsf{\upshape{\scriptsize{USE}}}\xspace}
\def\return{\textsf{\upshape{\scriptsize{RET}}}\xspace}
\def\killed{\textsf{\upshape{\scriptsize{DEF}}}\xspace}
\def\stale{\textsf{\upshape{\scriptsize{STALE}}}\xspace}

\def\ts#1{\textsf{\upshape{ts}(#1)}\xspace}

\newverbcommand{\tsv}{\textsf{\upshape{ts}}(}{)}
\def\parents#1{\textsf{\upshape{Par}(\code{#1})}\xspace}
\def\children#1{\textsf{\upshape{Chd}(\code{#1})}\xspace}

\declarecommand{\red}[1]{\textcolor{red}{#1}}
\declarecommand{\blue}[1]{\textcolor{blue}{#1}}
\declarecommand{\cyan}[1]{\textcolor{cyan}{#1}}
\declarecommand{\green}[1]{\textcolor{green}{#1}}
\declarecommand{\gray}[1]{\textcolor{gray}{#1}}
\declarecommand{\darkgreen}[1]{\textcolor{OliveGreen}{#1}}
\declarecommand{\pinegreen}[1]{\textcolor{pinegreen}{#1}}
\declarecommand{\purple}[1]{\textcolor{RoyalPurple}{#1}}

\ifnotes
\declarecommand{\agp}[1]{\textcolor{blue}{[Aditya: #1]}}
\declarecommand{\aditya}[1]{\agp{#1}}
\declarecommand{\agpquote}[2]{\textcolor{blue}{\string{#1\string}[Aditya: #2]}}
\declarecommand{\agpres}[1]{\ignorespaces}
\declarecommand{\smacke}[1]{\textcolor{cyan}{[Stephen: #1]}}
\declarecommand{\smackeres}[1]{\ignorespaces}
\declarecommand{\dlee}[1]{\textcolor{purple}{[Doris L: #1]}}
\declarecommand{\andrew}[1]{\textcolor{brown}{[Andrew: #1]}}
\declarecommand{\ronitt}[1]{\textcolor{pinegreen}{[Ronitt: #1]}}
\declarecommand{\agpst}[2]{\textcolor{blue}{[Aditya: \sout{#1} #2]}}
\declarecommand{\smackout}[2]{\textcolor{red}{\sout{#1} #2}}
\declarecommand{\smackoutres}[2]{\red{#2}}
\declarecommand{\resolved}[1]{\textcolor{green}{[Stephen: #1]}}
\declarecommand{\resolvedres}[1]{\ignorespaces}
\declarecommand{\strike}[1]{\sout{#1}}

\else
\declarecommand{\agp}[1]{\ignorespaces}
\declarecommand{\aditya}[1]{\agp{#1}}
\declarecommand{\agpquote}[2]{#1}
\declarecommand{\agpres}[1]{\ignorespaces}
\declarecommand{\smacke}[1]{\ignorespaces}
\declarecommand{\smackeres}[1]{\ignorespaces}
\declarecommand{\dlee}[1]{\ignorespaces}
\declarecommand{\andrew}[1]{\ignorespaces}
\declarecommand{\ronitt}[1]{\ignorespaces}
\declarecommand{\agpst}[2]{#2}
\declarecommand{\smackout}[2]{\red{#2}}
\declarecommand{\smackoutres}[2]{\red{#2}}
\declarecommand{\resolved}[1]{\ignorespaces}
\declarecommand{\resolvedres}[1]{\ignorespaces}
\declarecommand{\strike}[1]{\ignorespaces}

\fi

\ifcamera
\newcommand{\todo}[1]{#1}
\else
\newcommand{\todo}[1]{\red{#1}}
\fi

\ifarxiv
\newcommand{\figs}{.}
\else
\newcommand{\figs}{./figs}
\fi

\def\papertitle{Fine-Grained Lineage for Safer Notebook Interactions\xspace}
\def\paperauthors{Stephen Macke, Hongpu Gong, Doris Jung-Lin Lee, Andrew Head, Doris Xin, and Aditya Parameswaran}
\ifvldb
\fi

\ifvldb
\else
\newcommand\vldbvolume{14}
\newcommand\vldbissue{6}
\newcommand\vldbyear{2021}
\newcommand\vldbdoi{10.14778/3447689.3447712}
\newcommand\vldbpages{1093-1101}
\newcommand\vldbavailabilityurl{https://github.com/nbsafety-project/}
\newcommand\vldbpagestyle{empty}
\newcommand\vldbauthors{\paperauthors}
\newcommand\vldbtitle{\shorttitle} 
\fi

\def\root{.}
\ifarxiv
\def\figs{.}
\else
\def\figs{./figs}
\fi

\ifvldb
\title{\papertitle\\\textmd{\Large \papertext{[Scalable Data Science Track]}\techreport{[Technical Report]}}}
\author{%
\alignauthor Stephen Macke$^{1,2}$~~~~Hongpu Gong$^2$~~~~Doris Jung-Lin Lee$^2$~~~~Andrew Head$^2$\\Doris Xin$^2$~~~~~Aditya Parameswaran$^2$\\
       \affaddr{\scalebox{1.0}{$^1$University of Illinois (UIUC) \quad\quad $^2$UC Berkeley}} \\
       \affaddr{\scalebox{1.0}{\string{smacke,ruiduoray,dorislee,andrewhead,dorx,adityagp\string}@berkeley.edu}}%
}
\fi

\colorlet{BLUE}{blue}
\declarecommand{\rev}[1]{#1}
\declarecommand{\sectionrev}[1]{#1}
\newverbcommand{\rverb}{}{}

\begin{document}
\ifrebuttal
\pagenumbering{roman}
\input{rebuttal.tex}
\newpage
\pagenumbering{arabic}
\fi

\ifvldb
\maketitle
\else
\title{Fine-Grained Lineage for Safer Notebook Interactions}
\author{%
Stephen Macke,$^{1,2} \quad$ Hongpu Gong,$^2 \quad$ Doris Jung-Lin Lee,$^2 \quad$ Andrew Head,$^2$
}
\author{%
Doris Xin,$^2 \quad$ Aditya Parameswaran$^2$
}
\affiliation{%
\institution{\scalebox{1.0}{$^1$University of Illinois (UIUC)}}
\institution{\scalebox{1.0}{$^2$University of California, Berkeley}}
\institution{\scalebox{1.0}{\string{smacke,ruiduoray,dorislee,andrewhead,dorx,adityagp\string}@berkeley.edu}}
}
%
%
%
%
%


\fi

\begin{abstract}
Computational notebooks have
emerged as the platform of choice for data science
and analytical workflows,
enabling rapid iteration and exploration.
By keeping intermediate program state in memory and segmenting
units of execution into so-called ``cells'',
notebooks allow users to 
\techreport{execute their workflows interactively and}
enjoy particularly tight feedback. However, as cells are added,
removed, reordered, and rerun, this hidden intermediate state
accumulates\techreport{ in a way that is not necessarily correlated with
the notebook's visible code},
making execution behavior difficult to reason about,
and leading to errors and lack of reproducibility.
We present \name, a custom
Jupyter kernel that uses runtime tracing
and static analysis to automatically
manage lineage associated with cell execution
and global notebook state.
\name detects and prevents errors that
users make during unaided notebook interactions,
all while preserving the flexibility of existing notebook semantics.
We evaluate \name's ability to
prevent erroneous interactions
by replaying and analyzing \numworking real notebook sessions.
Of these, \name
identified \numunsafe sessions with potential safety errors,
and in the remaining \numsafe sessions,
the cells that \name identified as resolving safety issues
were more than $7\times$ more likely to be selected by users
for re-execution
compared to a random baseline,
even though the users were not using \name
and were therefore not influenced by its suggestions.




\end{abstract}

\iftech
\else
\maketitle

\pagestyle{\vldbpagestyle}
\begingroup
\markeverypar{\the\everypar\looseness=0} 
\small\noindent\raggedright\textbf{PVLDB Reference Format:}\\
\vldbauthors. \vldbtitle. PVLDB, \vldbvolume(\vldbissue): \vldbpages, \vldbyear.\\
\href{https://doi.org/\vldbdoi}{doi:\vldbdoi}
\endgroup
\begingroup
\markeverypar{\the\everypar\looseness=0} 
\renewcommand\thefootnote{}\footnote{\noindent
This work is licensed under the Creative Commons BY-NC-ND 4.0 International License. Visit \url{https://creativecommons.org/licenses/by-nc-nd/4.0/} to view a copy of this license. For any use beyond those covered by this license, obtain permission by emailing \href{mailto:info@vldb.org}{info@vldb.org}. Copyright is held by the owner/author(s). Publication rights licensed to the VLDB Endowment. \\
\raggedright Proceedings of the VLDB Endowment, Vol. \vldbvolume, No. \vldbissue\ %
ISSN 2150-8097. \\
\href{https://doi.org/\vldbdoi}{doi:\vldbdoi} \\
}\addtocounter{footnote}{-1}\endgroup

\ifdefempty{\vldbavailabilityurl}{}{
\vspace{.3cm}
\begingroup\small\noindent\raggedright\textbf{PVLDB Artifact Availability:}\\
The source code, data, and/or other artifacts have been made available at \url{\vldbavailabilityurl}.
\endgroup
}

\fi

\renewcommand*\ttdefault{txtt}
\renewcommand*\sfdefault{cmss}

\section{Introduction}
\label{sec:intro:nb}

Computational notebooks
such as Jupyter~\cite{jupyter}
provide a flexible medium for developers, scientists,
and engineers to complete programming tasks interactively. Notebooks, like
simpler predecessor read-eval-print-loops (REPLs), do not terminate after executing, but
wait for the user to give additional instructions while keeping intermediate
programming state in memory.
Notebooks, however, are distinguished from REPLs by
\iftech
three key features:
\begin{denseenum}
\item The atomic unit of execution in notebooks is the {\em cell}, composed
	  of a sequence of one or more programming statements, rather than a single
	  programming statement;
\item Notebooks allow users to easily refer back to previous cells to make
	  edits and potentially re-execute; and
\item Notebooks typically allow code and documentation to be interspersed, following
	  the {\em literate programming}~\cite{knuth1984literate} paradigm.
\end{denseenum}
\else
their
use of
the {\em cell} as the atomic
unit of execution, allowing users to edit and re-execute
previous cells\techreport{, as well as intersperse code with
documentation, following the
{\em literate programming}~\cite{knuth1984literate} paradigm}.
This cell-based iterative execution modality is a particularly good
fit for the exploratory, ad-hoc nature of modern data science.
\fi

As a result, the
IPython Notebook project~\cite{shen2014interactive}, and its successor,
Project Jupyter~\cite{jupyter},
have both grown rapidly in popularity.
\iftech
These projects decouple
the server-side {\em kernel} (responsible for running user code) from a browser-accessible
client side (providing the user interface).
Since Jupyter uses familiar web technologies to implement the layer that facilitates communication between the UI and the kernel,
it crucially allows users to run notebooks on any platform. Furthermore, Jupyter's
decoupled architecture allows users to leverage powerful server computing
resources from modest client-side hardware, particularly useful
for coping with the ever-increasing scale of modern dataset sizes.

These key features, along with the natural ergonomics of interactive computing
offered by the notebook interface, have led to an explosion in Jupyter's usage.
\fi
With more than $4.7$ million notebooks on GitHub as of March 2019~\cite{yan2020auto}\techreport{ and with hosted solutions
offered by a plethora of data analytics companies}, Jupyter has been
called ``data scientists' computational notebook of choice''~\cite{perkel2018jupyter}\techreport{,
and was recognized by the ACM Software System award in 2018~\cite{acmSoftwarePressRelease}}.
We focus on Jupyter here due to its popularity,
but
\ifvldb
we note that
\fi
our ideas are applicable to computational
notebooks in general.

Despite the tighter feedback enjoyed by users of computational notebooks, and,
in particular, by users of Jupyter, notebooks have a number of drawbacks when used
for more interactive and exploratory analysis.
Compared to conventional programming environments,
interactions such as
{\em out-of-order cell execution},
{\em cell deletion}, and
{\em cell editing and re-execution}
can all complicate the relationship between the code visible on screen and the
resident notebook state.
Managing interactions with this hidden notebook state is thus a burden shouldered by users,
who must remember what they have done in the past\techreport{,
since these past interactions cannot in general be reconstructed
from what is on the screen}.\dlee{Say something about managing hidden state in-the-head is a cognitively demanding activity on top of an already complex task of exploratory programming. So automating this away would be helpful.}

\dlee{Up to this point, it's not entirely clear as an outsider why notebooks allow you to do these weird execution behaviors. We should argue that the exploratory programming and notebook paradigm affords exploration (and is here to stay), so the solution is not to get rid of this (i.e., go back to file + script mode), but some type of safeguarding on top of this interactive paradigm.}

\begin{figure}[t]
\begin{lstjupyter}[cellnum={[1]},cellnumoffs=\thesissub{-7.5mm}{-29mm}]{In1}
 def !\HB!custom_agg!\HE!(series):
     ...
\end{lstjupyter}
\makesecondcellmarker{In1}{4}

\begin{lstjupyter}[cellnum={[2]},cellnumoffs=\thesissub{-7.5mm}{-29mm}]{In2}
 !\HB!agg_by_col!\HE! = {'A': 'min', 'B': !\HB!custom_agg!\HE!}
\end{lstjupyter}

\begin{lstjupyter}[cellnum={[3]},cellnumoffs=\thesissub{-7.5mm}{-29mm}]{In3}
 !\HB!df_x_agg!\HE! = df_x.agg(!\HB!agg_by_col!\HE!)
 !\HB!df_y_agg!\HE! = df_y.agg(!\HB!agg_by_col!\HE!)
\end{lstjupyter}
\makesecondcellmarker{In3}{5}
\ifvldb
\else

\vspace{-1.5em}
\fi
\begin{tikzpicture}[remember picture, overlay]
\tikzset{myarrow/.style={->, >=latex', shorten >=1pt, thick}}
\draw[myarrow,->] (In1-cell-marker.south) to [out=90,in=90] (In2-cell-marker.north);
\draw[myarrow,->] (In2-cell-marker.south) to [out=90,in=90] (In3-cell-marker.north);
\draw[myarrow,->] ([shift={(1mm,0)}]In3-cell-marker.north) to [out=45,in=-90] ([shift={(-0.5mm,0)}]In1-second-marker.south);
\draw[myarrow,->] ([shift={(1mm,0)}]In1-second-marker.south) to ([shift={(1mm,0)}]In3-second-marker.north);
\connectzm{In1-1}{blue}
\connectzm{In2-1}{red}
\connectzm{In2-2}{red}
\connectzm{In3-1}{blue}
\connectzm{In3-2}{red}
\connectzm{In3-3}{blue}
\connectzm{In3-4}{red}
\end{tikzpicture}
\ifthesis
\vspace{2em}
\else
\vspace{1em}
\fi
\ifvldb
\caption{Example sequence of notebook interactions leading to a stale symbol usage.
Symbols with timestamps
\blue{$\leq 3$} are shown with a \blue{blue} border, while symbols
with timestamps \red{$>3$} are shown with a \red{red} border.
\dlee{The Figure makes sense after reading the Case Study paragraph, but doesn't really make sense at a glance when standing on its own. Can you color the cell border to be similar to what nbsafety does to show which cell is stale. For someone who is not familiar with Jupyter, they might not know that "In[...]" means execution order. If it would help, you can separate the diagram into two snapshots (state before staleness appears, and state after), to make the history more linear} \andrew{Agreed with Doris, this figure was a little tricky for me to parse}}
\else
\caption{Example sequence of notebook interactions leading to a stale symbol usage.
Symbols with timestamps
\blue{$\leq 3$} (resp.\@ \red{$>3$}) are shown with \blue{blue} (resp.\@ \red{red}) borders.}
\fi
\label{fig:intro-interactions}
\papertext{\vspace{-.1em}}
\end{figure}
\topic{Illustration}
Consider the sequence of notebook interactions depicted in \Cref{fig:intro-interactions}.
Each rectangular box indicates a cell,
the notebook's unit of execution.
The user first defines a custom aggregation function that,
along with \verb!min!, will
be applied to two dataframes, \verb!df_x! and \verb!df_y!,
and executes it as cell \verb![1]!.
Since both aggregations will be applied to both dataframes, the user
next gathers them into a function dictionary in the second cell
(executed as cell \verb![2]!).
After running the third cell,
which corresponds to applying the aggregates to \verb!df_x! and \verb!df_y!,
the user realizes an error in the logic
of \verb!custom_agg! and goes back to the first cell to fix the bug.
They re-execute this cell after making their update,
indicated as \verb![4]!.
However, they forget that the old version of \verb!custom_agg! still
lingers in the \verb!agg_by_col! dictionary
and rerun the third cell (indicated as \verb![5]!) without rerunning the second cell.
We deem this an {\em unsafe} execution,
because the user intended for the change to \verb!agg_by_col! to be reflected
in \verb!df_agg_x! and \verb!df_agg_y!, but it was not.
Upon inspecting the resulting dataframes \verb!df_x_agg! and \verb!df_y_agg!,
the user may or may not realize the error.
In the best case,
user may identify the error and rerun the second cell.
In the worst case,
users may be deceived into thinking that their
change had no effect\techreport{, with the
original error then propagating throughout
the notebook}.

This example underscores the inherent difficulty in manually managing
notebook state, inspiring colorful criticisms such as a talk titled
``I Don't Like Notebooks'' presented at JupyterCon 2018~\cite{idontlikenotebooks}.
In addition to the frustration that users experience when
spending valuable time debugging state-related errors,
such bugs can lead to invalid research results and 
hinder reproducibility\techreport{,
inspiring the claim that one must
``restart and run all or it didn’t happen''~\cite{perkel2018jupyter}
if presenting results via a notebook medium}.

\topic{Key Research Challenges}
The goal of this \work is to
{\em \ul{develop techniques to automatically identify and prevent
potentially unsafe cell executions}},
without sacrificing existing familiar notebook semantics.
We encounter 
a number of challenges toward this end:


\emtitle{1. Automatically detecting unsafe interactions.}
To detect unsafe interactions due to symbol staleness issues, \techreport{the approach that immediately suggests itself is static code analysis, but upon
deeper reflection,} it becomes clear that static analysis on its own is not enough.
A static approach must necessarily be overly conservative when
gathering lineage metadata / inferring dependencies,
as it must consider all branches of control flow.
On the other hand, {\em some} amount of static analysis is necessary
so that users can be warned before they execute an unsafe cell (as opposed to during cell execution,
by which time the damage may already be done);
finding the right balance is nontrivial.

\emtitle{2. Automatically resolving unsafe behavior with suggested fixes.}
In addition to detecting potentially unsafe interactions, we should ideally
also identify which cells to run in order to resolve staleness issues.
A simpler approach may be to automatically rerun cells when a potential
staleness issue is detected (as in Dataflow notebooks~\cite{koop2017dataflow}),
but in a flexible notebook environment, there could potentially
be more than one cell whose re-executions would all resolve a particular staleness issue;
identifying these to present them as options to the user requires
a significant amount of nontrivial static analysis.



\emtitle{3. Maintaining interactive levels of performance.}
\rev{We must address the aforementioned challenges} 
without introducing unacceptable latencies
or memory usage.
First, we must ensure that any lineage metadata we introduce
does not grow too large in size.
Second, efficiently identifying cells that resolve staleness issues is also nontrivial.
Suppose we are able to detect cells with staleness issues,
and we have detected such issues in cell $c_s$.
We can check whether prepending some cell $c_r$
(and thereby executing $c_r$ first before $c_s$)
would fix the staleness issue (by, \eg, detecting whether
the merged cell $c_r\oplus c_s$ has staleness issues),
but we show in \Cref{sec:efficient-refresher} that a direct implementation of this idea
scales quadratically in the number of cells in the notebook\techreport{
and therefore quickly loses viability}.
\techreport{Developing an efficient approach is thus a major challenge.}

\vspace{2.5pt}
Despite previous attempts to
address these challenges and to facilitate safer interactions
with global notebook state~\cite{koop2017dataflow,nodebook,datalore},
to our knowledge,
\name is the first to do so while preserving 
\techreport{the flexibility of} 
existing notebook semantics.
For example, Dataflow notebooks~\cite{koop2017dataflow}
require users to explicitly annotate cells with their dependencies,
and force
the re-execution of cells whose dependencies have changed.
Nodebook~\cite{nodebook} \rev{and the Datalore kernel~\cite{datalore}} attempt
to enforce a temporal ordering of variable definitions in the order that
cells appear, again forcing users to compromise on flexibility.
In the design space of computational notebooks~\cite{LauVLHCC2020}, Dataflow notebooks
observe {\em reactive} execution order, 
while Nodebook \rev{and Datalore's kernel} observe {\em forced in-order} execution.
However, a solution that preserves {\em any-order} execution semantics,
while simultaneously helping users
avoid errors that are only made possible due to such flexibility,
has heretofore evaded development.

%
%
%
%

\topic{Contributions}
To address these challenges, we develop \name, a custom Jupyter kernel and frontend
for automatically detecting unsafe interactions and alerting users,
all while maintaining interactive
levels of performance
and preserving existing notebook semantics.
\iftech
Installing \name is easy:
after running a single installation command\footnote{\code{pip install nbsafety}},
\else
\rev{After a single installation command~\cite{nbsafety-code},}
\fi
users of both JupyterLab and traditional Jupyter notebooks
can opt to use the \name kernel as a drop-in
replacement for Jupyter's built-in Python 3 kernel.
\name introduces two
key innovations 
to address the challenges outlined above:


\emtitle{1. Efficient and accurate detection of staleness issues in cells via novel joint dynamic and static analysis.}
The \name kernel combines runtime tracing with static analysis
in order to detect and prevent notebook interactions
that are unsafe due to staleness issues of the form
seen in \Cref{fig:intro-interactions}.
The tracer (\S\ref{sec:lineage:nb})
instruments each program statement so that program
variable definitions
are annotated with parent
dependencies and cell execution timestamps.
This metadata is then used by a {\em runtime state-aware static checker} (\S\ref{sec:analysis:nb})
that combines said metadata with static program analysis techniques to determine whether
any staleness issues are present {\em prior} to the start of cell execution.
This allows \name to present users with {\em cell highlights} (\S\ref{sec:highlights:nb}) that warn them about cells that are
unsafe to execute due to staleness issues
{\em before} they try executing such cells,
thus preserving desirable atomicity of cell executions
present in traditional notebooks.

\emtitle{2. Efficient resolution of staleness issues.}
Beyond simply detecting staleness issues, we also show how
to detect cells whose re-execution would resolve such staleness issues 
--- but doing so {\em efficiently}
required us to \rev{leverage a lesser-known} analysis technique called {\em \rev{initialized variable} analysis}
\iftech
(\S\ref{sec:inverse-liveness})
\else
(\S\ref{sec:analysis:nb})
\fi
tailored to this use case. We show how \rev{initialized} analysis
brings staleness resolution complexity down from time quadratic in the number of cells in the notebook
to linear, crucial for large notebooks.

\vspace{2.5pt}
We validate our design choices for \name by replaying and analyzing of a corpus of
\numworking execution logs of real notebook sessions, scraped from GitHub (\S\ref{sec:experiments:nb}).
In doing so, \name identified that \numunsafe sessions had potential safety
errors, and upon sampling these for manual inspection, we found several
with particularly egregious examples of confusion and wasted effort by
real users that would have been saved with \name. 
After analyzing the \numsafe remaining sessions,
we found that cells suggested by \name as resolving staleness issues were strongly favored by users for re-execution---more than $7\times$ more likely to be selected compared to random cells,
even though these user interactions 
were originally performed without \name and
therefore were not influenced by its suggestions.
Overall, our empirical study indicates that
\name can reduce cognitive overhead associated with manual maintenance of global notebook state
under 
any-order execution semantics, and in doing so,
allows users to focus their efforts more on 
\techreport{the already challenging task of } 
exploratory data analysis,
and less on avoiding and fixing state-related notebook bugs.
\andrew{I'm not sure I fully understand why this indicates a success for nbsafety. It might need to be spelled out a bit more.}
\aditya{I agree, this has to be framed as ``relieving
users of the burden of having to figure out what to re-execute
by doing it for them. Moreover it identified a number of
staleness issues.''}
\smacke{See new sales pitch above.}

Our free and open source code is available publicly on GitHub~\cite{nbsafety-code}.


\techreport{\topic{Organization}
\Cref{sec:arch:nb} gives a high-level overview of \name's architecture and how it
integrates into a notebook workflow. The next three sections drill into each
of \name's components: \Cref{sec:lineage:nb} describes how the tracer maintains lineage metadata,
\Cref{sec:analysis:nb} describes the static analyses employed by the checker,
and \Cref{sec:highlights:nb} describes how these two components feed into the frontend
in order help users avoid and resolve safety issues.
We empirically validate \name's ability to highlight (i) cells that should likely be avoided
and (ii) cells that should likely be re-executed in
\iftech
\Cref{sec:experiments:nb} before surveying related work and concluding (\S\ref{sec:related:nb}, \S\ref{sec:conclusion:nb}).
\else
\Cref{sec:experiments:nb}.
\fi
}

\section{Architecture Overview}
\label{sec:arch:nb}


In this section, we give an overview of \name's components
and how they integrate into the notebook workflow.


\techreport{We now describe the operation of each component
in more detail.}

\input{\root/arch-lifecycle.tex}

\topic{Overview}
\name integrates into a notebook workflow
according to \Cref{fig:nblife:nb}.
As depicted, all components of \name are invoked
upon each and every cell execution.
When the user submits a request to run a cell,
the tracer (\bcircle{1}, \S\ref{sec:lineage:nb})
instruments the executed cell,
updating lineage metadata associated with each variable
as each line executes. Once the cell
finishes execution,
the checker (\bcircle{2}, \S\ref{sec:analysis:nb})
performs {\em liveness analysis}~\cite{aho1986compilers}
\techreport{(a standard technique for finding non-redefined symbols in a program)}
and {\em \rev{initialized variable} analysis}~\cite{moller2012static}
\techreport{(discussed in \S\ref{sec:inverse-liveness})}
for every cell in the notebook.
By combining the results of these analyses
with the lineage metadata computed
by the tracer, the frontend
(\bcircle{3}, \S\ref{sec:highlights:nb}) is able to highlight
cells that are unsafe due to staleness issues
of the form seen in \Cref{fig:intro-interactions},
as well as cells that resolve such staleness issues.

\topic{\bcircle{1} Tracer}
The \name tracer maintains dataflow dependencies for each symbol that
appears in the notebook in the form of lineage metadata.
It leverages Python's built-in tracing capabilities~\cite{python-tracing},
which allows it to run custom code upon four different kinds of events: (i) \verb!line! events,
when a line starts to execute;
(ii) \verb!call! events, when a
function is called,
(iii) \verb!return! events, when a function returns, and
(iv) \verb!exception! events, when an exception occurs.

To illustrate its operation,
consider that,
the first time $c_3$ in \Cref{fig:intro-interactions} is executed, symbols \verb!df_agg_x! and \verb!df_agg_y! are undefined.
Before the first line runs, a \verb!line! event occurs, thereby trapping
\aditya{?}\smacke{this is a standard systems programming term. often refers to the
equivalent of an exception handler,
but wikipedia says it's OK to use it as a more general callback handler:
``In some usages, the term trap refers specifically
to an interrupt intended to initiate a context switch to a monitor program or debugger.''}
into the tracer.
The tracer has access to the line of code that triggered the \verb!line! event and parses it as an \verb!Assign! statement
in Python's grammar, followed by
a quick static analysis to determine that the symbols \verb!df_x! and \verb!agg_by_col! appear
on the right hand side of the assignment (\ie, these symbols appear in \use[\rhs of the \verb!Assign!]).
Thus, these two will be the dependencies for
symbol \verb!df_agg_x!. Since $c_3$ is the third cell executed, the tracer furthermore
gives \verb!df_agg_x! a timestamp of $3$.
Similar statements hold for \verb!df_agg_y! once the second line executes.

\topic{\bcircle{2} Checker}
The \name static checker performs two kinds of program analysis:
(i) {\em liveness analysis}, and (ii) {\em \rev{initialized variable analysis}}.
The \name liveness checker helps to detect safety issues by determining
which cells have live references to stale symbols.
For example, in \Cref{fig:intro-interactions},
\verb!agg_by_col!, which is stale, is live in $c_3$---this information can be used
to warn the user before they execute $c_3$.
Furthermore,
the \rev{initialized} checker serves as a key component
for efficiently computing resolutions to staleness issues,
as we later show in \Cref{sec:analysis:nb,sec:highlights:nb}.

\begin{figure}[t]
\begin{center}
\includegraphics[width=\thesissub{0.75\linewidth}{0.75\linewidth}]{\figs/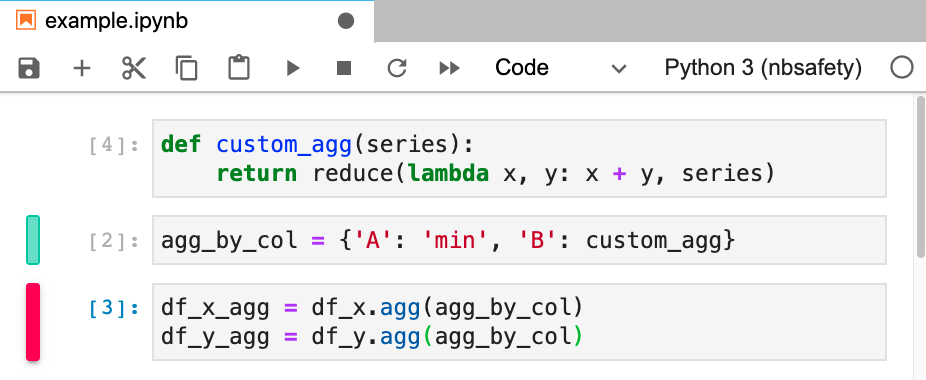}
\ifvldb
\caption{\name highlights cells that are unsafe to execute with \unsafehl{staleness warnings}
and cells that resolve such issues with \refresherhl{cleanup suggestions}.}
\else
\caption{\name highlights unsafe cells with \unsafehl{staleness warnings}
and cells that resolve staleness issues with \refresherhl{cleanup suggestions}.}
\fi
\label{fig:highlight-example}
\end{center}
\end{figure}

\topic{\bcircle{3} Frontend}
The \name frontend uses the results of the static checker
to highlight cells of interest. For example, in \Cref{fig:highlight-example},
which depicts the original example from \Cref{fig:intro-interactions} (but before
the user submits $c_3$ for re-execution), $c_3$ is given a \unsafehl{staleness warning} highlight
to warn the user that re-execution could have incorrect behavior due to staleness issues.
At the same time, $c_2$ is given a \refresherhl{cleanup suggestion}
highlight,
because rerunning it would resolve the staleness 
in $c_3$.
The user can then leverage the extra visual cues to make a more informed decision
about which cell to next execute\techreport{,
possibly preferring to execute cells that resolve staleness
issues before cells with staleness issues}.

\iftech
\topic{\rev{Design Philosophy}}
\rev{In contrast with systems that automatically resolve staleness,
such as Datalore~\cite{datalore} or Nodebook~\cite{nodebook},
\name is designed to
be as non-intrusive as possible, while providing useful information.
\name thus only attempts to make a ``passive observer'' guarantee to ensure that existing notebook semantics are
preserved, via tracing and static analyses that monitor, and do not alter, notebook behavior.
We make this decision partially because program analysis in a dynamic language
like Python is notoriously difficult; as such, \name's analysis components are
unable to make any formal guarantees regarding soundness or completeness,
instead opting for a ``best-effort'' attempt to provide the user with useful information.}

\rev{Although the guarantee \name offers regarding preservation of semantics may seem weak at first glance,
we show later (\S\ref{sec:comparison:nb}) that both Datalore and Nodebook crash on reasonable programs due to their intrusiveness.
Furthermore, we also show how existing semantics lend themselves to multiverse analyses~\cite{steegen2016increasing},
and point to specific examples of such analyses in our experiments.
Finally, \name is not necessarily intended to
be a substitute for systems like Dataflow
notebooks~\cite{koop2017dataflow} that allow users to give names to cell outputs
and explicitly reference them; the two techniques can complement each other, as
the issues faced by notebook users do not automatically disappear by using
Dataflow notebooks.}
\fi

Overall, each of \name's three key components play crucial roles
in helping users avoid and
resolve unsafe interactions due to staleness issues
\rev{without compromising existing notebook program semantics.}
We describe each component
\ifvldb
in detail
\else
\fi
in the following sections.

\section{Lineage Tracking}
\label{sec:lineage:nb}

In this section, we describe how \name traces cell execution
in order to maintain symbol lineage metadata,
and how such metadata aids in the detection and resolution of
staleness issues.
\techreport{We begin by introducing helpful terminology and
formalizing our notion of staleness beyond the
intuition we gave \techreport{in \Cref{fig:intro-interactions} of \Cref{sec:intro:nb}.}
\papertext{from the example in \Cref{fig:intro-interactions}.}}

\subsection{Preliminaries}
\label{sec:lineage-prelim:nb}

We begin defining our use of the term {\em symbol}.
\begin{definition}[Symbol]
\label{def:sym:nb}
A {\em symbol} is any piece
of data in notebook scope that can be referenced by a (possibly qualified) name.
\end{definition}
For example, if \verb!lst! is a list with 10 entries, then \verb!lst!, \verb!lst[0]!,
and \verb!lst[8]! are all symbols. Similarly, if \verb!df! is a dataframe with a column
named ``col'', then \verb!df! and \verb!df.col! are both symbols.
Symbols can be thought of as a generalized notion of variables
that allow us treat different nameable objects in Python's
data model in a unified manner.

\name augments each symbol with additional lineage metadata in the form of {\em timestamps}
and {\em dependencies}.

\begin{definition}[Timestamp]
\label{def:ts:nb}
A symbol's {\em timestamp}
is the execution counter of the cell that most recently modified that symbol.
Likewise, a cell's timestamp is the execution counter corresponding to the
most recent time that cell was executed.
\end{definition}
\aditya{You need to define / provide intuition for exec counter.}
For a symbol \verb!s! or a cell $c$, we denote its timestamp as \ts{s} or \ts{$c$}, respectively.
For example, letting  $c_1$, $c_2$, and $c_3$ denote the three cells in \Cref{fig:intro-interactions},
we have that \tsv!custom_agg! $= \ts{$c_1$} = 4$, since \verb!custom_agg! is last updated in $c_1$,
which was executed at time $4$.

\begin{definition}[Dependencies]
\label{def:dep:nb}
The {\em dependencies} of symbol \verb!s! are those symbols
that contributed to
\ifvldb
the computation of \verb!s!
\else
\verb!s!'s computation
\fi
via direct dataflow.
\end{definition}
In \Cref{fig:intro-interactions}, \verb!agg_by_col! depends on \verb!custom_agg!,
while\\ \verb!df_x_agg! depends on \verb!df_x! and \verb!custom_agg!.
\ifvldb
We write \parents{s} to denote the dependencies of \verb!s!.
\else
We denote the dependencies of \verb!s! with \parents{s}.
\fi
\iftech
Similarly, if $\verb!t!\in\parents{s}$, then $\verb!s!\in\children{t}$.
\fi

A major contribution of \name is to highlight cells with unsafe usages
of {\em stale symbols}, which we define recursively as follows:
\begin{definition}[Stale symbols]
\label{def:stalesym:nb}
A symbol \verb!s! is called {\em stale}
if there exists some
\verb!s!$'\in\parents{s}$
such that
$\ts{s$'$} > \ts{s}$,
or \verb!s!$'$ is itself stale;
that is, \verb!s! has a parent that is either itself stale or more up-to-date than \verb!s!.
\end{definition}
In \Cref{fig:intro-interactions}, symbol \verb!agg_by_col! is stale, because \tsv!agg_by_col! $= 2$,
but \tsv!custom_agg! $= 4$. Staleness gives us a rigorous conceptual framework upon which to study
the intuitive notion that, because \verb!custom_agg! was updated,
we should also update its child \verb!agg_by_col! to prevent counterintuitive behavior.
\dlee{Perhaps you can split Figure 1 into two diagram, one is the showing the cell and explains the scenario. The other diagram is solely the DAG which you can put closer to this section to highlight parent child relationships. I would also untangle the graph a bit to have a more tree-like layout to illustrate your point.}

We now draw on these definitions as we
describe how \name maintains lineage metadata
\ifvldb
for symbols
\else
\fi
while tracing cell execution.

\subsection{Lineage Update Rules}
\label{sec:lineage-rules:nb}
\begin{table}[t]
\begin{densecenter}
{\papertext{\scriptsize}
\begin{tabular}{|c|c|c|c|c|c|}
\hline\rowcolor[HTML]{C5C5C5}{\bf AST Node} & {\bf Example} & {\bf Rule}\\ \hline
\textsf{Assign} & \code{a = $e$} & $\parents{a} = \use[e]$ \\ \hdashline
(target in RHS) & \code{a = a + $e$} & $\parents{a} = \parents{a} \cup \use[e]$ \\ \hdashline
\iftech
\rev{\textsf{Assign} with \textsf{Call}} & \rev{\code{a = f($e$)}} & \rev{$\parents{a} = \use[e] \cup \return[\code{f}]$} \\ \hdashline
\fi
\textsf{AugAssign} & \code{a += $e$} & $\parents{a} = \parents{a} \cup \use[e]$ \\ \hline
\textsf{For} & \code{for a in $e$:} & $\parents{a} = \use[e]$ \\ \hline
\textsf{FunctionDef} & \code{def f(a=$e$):} & $\parents{f} = \use[e]$ \\ \hline
\textsf{ClassDef} & \code{class c($e$):} & $\parents{c} = \use[e]$ \\ \hline
\end{tabular}}
\caption{\rev{Subset of l}ineage rules used by the \name tracer.}
\label{tab:lineage-rules}
\end{densecenter}
\ifvldb
\vspace{-10pt}
\fi
\end{table}

\name attempts to be non-intrusive when maintaining lineage
with respect to the Python objects that comprise the notebook's
state. To do so, we avoid modifying the Python objects created
by the user, instead creating ``shadow'' references to each symbol.
\name then takes a hybrid dynamic / static approach to updating
each symbol's lineage. After each
\ifvldb
Python
\fi
statement has finished
executing, the tracer inspects the AST node for the executed statement
and performs a lineage update according to the rules shown in \Cref{tab:lineage-rules}.

\topic{Example}
Suppose the statement
\[ \code{gen = map(lambda x: f(x), foo + [bar])} \]%
has just finished executing.
\rev{Using rule 1 of \Cref{tab:lineage-rules}, t}he
tracer will then statically analyze the right hand side in order to determine
\[ \text{\upshape \use[\code{map(lambda x: f(x), foo + [bar])}]} \]%
which is the set of used symbols that appear
in the RHS. In this case, the aforementioned set is $\{\verb!f, foo, bar!\}$ --- everything else is either
a Python built-in (\verb!map!, \verb!lambda!),
or an unbound symbol (\ie in the case of the lambda argument \verb!x!).
The tracer will thus set
\iftech
\[ \parents{gen} = \{\verb!f!, \verb!foo!, \verb!bar!\} \]
\else
$\parents{gen} = \{\verb!f!, \verb!foo!, \verb!bar!\}$
\fi
and will also
\iftech
set \ts{gen} to the current cell's execution counter.
\else
update \ts{gen}.
\fi
\dlee{I didn't quite follow this example, perhaps I was a bit tripped up by `USE'. It was unclear if this is just illustrating one of the rules in Table 1 or all of the rules in Table 1.}

\iftech
\else
\topic{\rev{Fine-Grained Lineage for Attributes and Subscripts}}
\rev{\name is able to track lineage at a finer granularity than
simply top-level symbols. For example, \name tracks parents and
children of subscript symbols like} \rverb!x[0]!
\rev{and attribute symbols like} \rverb!x.a! \rev{(as well as combinations thereof)
as first-class citizens, in addition to those of top-level symbols such as}
\rverb!x!\rev{. Please see the technical report~\cite{nbtech} for more details.}
\fi

\iftech
\topic{\rev{Handling Function Calls and Returns}}
\rev{Recall that, in Python, a function may ``capture'' symbols
defined in external scope by referencing them. In this case,
the lineage update rule for a function call needs to be aware
of the symbols referenced by the function's return
statement. As such, the \name tracer saves these symbols
when it encounters a return statement, and loads them upon
encountering a} \rverb!return! \rev{event, so that they are
available for use with lineage update rules, \eg, the third
entry of \Cref{tab:lineage-rules}.}
\fi

\iftech
\topic{Rationale for Tracing}
\smacke{Candidate for techreport}
\aditya{I like this but can be TR'd}
Given that the tracer is already performing some static analysis
as each Python statement executes, a natural question is: why
should we trace cell execution at all, instead of performing
static analysis on an entire cell in order to make
lineage updates? The answer is that, when a cell executes, a relatively
small number of control flow paths may be taken, and a purely
static approach must consider them all in order to be conservative. This
may cause each \parents{s} to be much larger than necessary, or to be
unnecessarily overwritten if, \eg, \verb!s! is assigned in some unexecuted
control flow path. This can happen due to, \eg, untaken branches, or if
an exception is thrown mid-cell.
Indeed, we attempted to infer lineage updates statically
in an earlier version of \name, but found this to be too coarse-grained in order
to derive real benefits.
\fi

\topic{\rev{Staleness Propagation}}
We already saw that the tracer annotates each symbol's shadow reference
with timestamp and lineage metadata. Additionally, it tracks
whether each symbol is stale, as this cannot be inferred solely
from timestamp and lineage metadata. To see why, recall
%
the definition of staleness: a symbol \verb!s! is stale
if it has a more up-to-date parent
(\ie, an $\verb!s!'\in\parents{s}$ with $\ts{s$'$} > \ts{s})$,
{\em or if it has a stale parent}, precluding the ability
to determine staleness locally.
Thus, when \verb!s! is updated, we perform a depth first search
starting from each child $\verb!c! \in \children{s}$ in order
to propagate the ``staleness'' to all descendants.

\iftech
\topic{\rev{Fine-Grained Lineage for Attributes and Subscripts}}
Recall that the \name tracer attempts to infer parent symbols
statically when making lineage updates. While this is possible
in many cases, there are limits to a purely static approach.
For example, the statement ``\verb!s = a.b().c!'' is valid Python code,
but, in general, it is impossible to statically determine what \verb!a.b()! returns.
As such, we must again rely on tracing to do so.
Unfortunately, Python's built-in tracing
abilities operate at the granularity of code lines, so that a \verb!return!
event does not tell us the point within a line to which control returns.
Thus, the tracer rewrites all
\textsf{Attribute} nodes in the statement's AST, so that the
earlier example will actually run as follows:
\[ \verb!s = trace(trace(a, 'b').b(), 'c').c! \]
The \verb!trace! function will first determine that the symbol name
\verb!b! is referenced within symbol \verb!a!'s namespace, setting the
current RHS parent symbol to \verb!a.b!. However, after the call event,
the tracer will update this to the \verb!c! symbol in the namespace of the return value
of \verb!a.b()!, so that the true parent symbol is pinpointed.

Subscripts are handled analogously to attributes.

\topic{Handling External Libraries}
\name assumes that imported libraries
do not have access to notebook state. Thus, when the tracer observes a call
into library code, it simply halts tracing, resuming once control returns
to the notebook.
\fi

\topic{Bounding Lineage Overhead}
\noindent Consider the following cell:
\begin{figure}[H]
\vspace{-0.5em}
\iftech
\begin{lstjupyter}[cellnum=,cellnumfill=none]{sc1}
  x = 0
  for i in random.sample(range(10**7), 10**5) + [42]:
      x += lst[i]
\end{lstjupyter}
\else
\begin{lstjupyter}[cellnum={[1]}]{sc1}
  x = 0
  for i in random.sample(range(10**7), 10**5):
      x += lst[i]
\end{lstjupyter}
\fi
\end{figure}
\vspace{1.5pt}
\noindent In order to maintain lineage metadata for symbol \verb!x! to $100\%$ correctness,
we would need to somehow indicate that \parents{x} contains
\iftech
\rverb!lst[i]! \rev{for all $10^5$ random indices} \rverb!i! \rev{(as well as for} \rverb!lst[42]!\rev{)}.
\else
\rverb!lst[i]! \rev{for all $10^5$ random indices} \rverb!i!.
\fi
It is impossible to maintain acceptable performance in general under these circumstances.
Potential workarounds include {\em conservative} approximations, as well
as {\em lossy} approximations. For example, as a conservative approximation,
we could instead specify that \verb!x! depends on \verb!lst!,
with the implication that it also depends on everything in \verb!lst!'s namespace.
However, this will cause \verb!x! to be incorrectly classified as stale
whenever \verb!lst! is mutated, \eg, if a new entry is appended.
\papertext{We therefore opted for a lossy approximation 
that we describe in our extended technical report~\cite{nbtech}.}
\iftech

Thus, \name makes a compromise and
sacrifices some correctness by {\em only instrumenting each AST statement the first time it executes}.
In this case, after the cell executes, \parents{x} will have a single entry: \verb!lst[0]!.
This helps to ensure that lineage metadata only grows in proportion to the amount of text
in the user's notebook.

Note that this approach can lead to some false negatives when determining which symbols are stale.
In the example above, if the user makes a point update to, \eg, \verb!lst[42]!, \verb!x! should
technically be considered stale, but will fail to be updated as so since
$\verb!lst[42]!\notin\parents{x}$. We consider this tradeoff worthwhile, as we found
that performance suffered greatly without it. \smacke{Forward ref to benchmark if time permits.}
\fi
\iftech

\topic{Garbage Collection}
Each symbol's shadow metadata maintains a weak reference to the symbol.
When a Python object is deleted, \eg, because the user executed a \verb!del! statement,
or because it was garbage collected by Python's built-in garbage collector, a callback associated
with the weak reference executes. This callback tells the metadata to delete itself
(including from any \parents{$\cdot$} or \children{$\cdot$} sets), thereby ensuring that lineage metadata
does not accumulate without bound over long notebook sessions.

\topic{\rev{Handling Aliases}}
\todo{Since multiple symbols can reference the same object in Python,
we need a mechanism to propagate mutations to some object to all
symbols that reference it. For example, if both x and y refer to
the same list, and then this list is appended to, both x and y should
be have their timestamps bumped. \name accomplishes this by creating
a registry that maps objects to all symbols referencing them.}
\else

\topic{\rev{Handling Calls to External Libraries}}
\rev{When \name's tracer traps due to a function call, it inspects the
location of the called function. If the called function was not defined
in the user's notebook, but in some imported file, \name disables tracing
until control returns to the notebook proper, since external files typically
do not have access to state defined in the notebook.
If an object in notebook state is passed
explicitly as, \eg, a function parameter, \name assumes the library does
not mutate it; we leave improvements to future work.
By disabling tracing when control is outside the notebook, we
ensure that additional tracing overhead is bounded by the size of the
user's notebook.}

\rev{Finally, our technical report~\cite{nbtech} contains additional
details surrounding the tracer, such as how we garbage collect shadow metadata,
how we handle mutations when variables alias each other, and}
how we handle namespaced symbols such as attributes
\techreport{ and subscripts (\eg, which would appear as \verb!a.b! or \verb!a[0]! in Python code)}.
\fi

\section{Liveness and \sectionrev{Initialized Analyses}}
\label{sec:analysis:nb}

\iftech
\thesissub{\begin{table*}}{\begin{figure}[t]}
\begin{tabular}{c|c}
\begin{minipage}{0.5\textwidth}
\begin{algo}[t]
\captionof{algocf}{Liveness checker}\label{alg:liveness:nb}
\algrenewcomment[1]{\texttt{/*} #1 \texttt{*/} \\}
\SetKwInOut{Input}{Input}
\SetKwInOut{Output}{Output}
\SetKwRepeat{Do}{do}{while}
\SetKwRepeat{Repeat}{repeat}{until}
\SetKwProg{myfunc}{function}{}{}
{\scriptsize
\Input{Cell $c$, CFG $G$ that stores successors $\textsf{succ}[s]$ for each \code{stmt} $s\in c$}
\Output{$\live(c)$}
\ForEach{\upshape\codespace{stmt}$s\in c$}{
	$\live_{out}[s] \gets \emptyset$\;
  {\color{linkpurple}\tikzmark{hl1-start}$\live_{in}[s] \gets \emptyset$\;}
	$\use[s] \gets \{\text{referenced symbols in $s$}\}$\;
	$\killed[s] \gets \{\text{assigned symbols or defined functions in $s$}\}$\;
}
{\color{linkpurple}$\live_{in}[G_{\code{exit}}] \gets \emptyset$\;}
\Repeat{
$\live_{*}[s] = \live'_{*}[s]\ \text{\upshape for both}\ *\in\{in,out\},\forall s\in c$}{%
	\ForEach{\upshape\codespace{stmt}$s\in c$}{
		$\live'_{out}[s] \gets \live_{out}[s]$\;
		$\live'_{in}[s] \gets \live_{in}[s]$\;
  {\color{linkpurple}
		\tikzmark{hl2-start}$\live_{out}[s]\gets\bigcup_{s'\in\textsf{succ}[s]}\live_{in}[s']$\;}
  {\color{linkpurple}
		$\live_{in}[s] \gets \use[s] \cup (\live_{out}[s] - \killed[s])$\;}
	}%
}
\tikzmark{hl3-start}\Return{\upshape $\live_{in}[G_\code{entry}]$}\;
}
\end{algo}

\end{minipage}
&
\begin{minipage}{0.5\textwidth}
\begin{algo}[t]
\captionof{algocf}{\rev{Initialized variable} checker}\label{alg:deadness:nb}
\algrenewcomment[1]{\texttt{/*} #1 \texttt{*/} \\}
\SetKwInOut{Input}{Input}
\SetKwInOut{Output}{Output}
\SetKwRepeat{Do}{do}{while}
\SetKwRepeat{Repeat}{repeat}{until}
\SetKwProg{myfunc}{function}{}{}
{\scriptsize
\Input{Cell $c$, CFG $G$ that stores predecessors $\textsf{pred}[s]$ for each \code{stmt} $s\in c$}
\Output{$\dead(c)$}
\ForEach{\upshape\codespace{stmt}$s\in c$}{
	$\dead_{in}[s] \gets \emptyset$\;
  {\color{linkpurple}$\dead_{out}[s] \gets \univ$\tikzmark{hl1-end}\;}\label{line:dead-init}
  $\use[s] \gets \{\text{referenced symbols in $s$}\}$\;
  $\killed[s] \gets \{\text{assigned symbols or defined functions in $s$}\}$\;
}
{\color{linkpurple}$\dead_{out}[G_{\code{entry}}] \gets \emptyset$\;}
\Repeat{\tikzmark{repeat-2}
$\dead_{*}[s] = \dead'_{*}[s]\ \text{\upshape for both}\ *\in\{in,out\},\forall s\in c$}{%
  \ForEach{\upshape\codespace{stmt}$s\in c$}{
  	$\dead'_{in}[s] \gets \dead_{in}[s]$\;
  	$\dead'_{out}[s] \gets \dead_{out}[s]$\;
  {\color{linkpurple}
  	$\dead_{in}[s]\gets\bigcap_{s'\in\textsf{pred}[s]}\dead_{out}[s']$\;}\label{line:dead-incoming}
  {\color{linkpurple}
    $\dead_{out}[s] \gets (\killed[s]-\use[s]) \cup \dead_{in}[s]$\tikzmark{hl2-end}}\;\label{line:dead-outgoing}
  }%
}
\Return{\upshape $\dead_{out}[G_\code{exit}]$}\tikzmark{hl3-end}\;
}
\end{algo}

\end{minipage}
\end{tabular}
\thesissub{\end{table*}}{\end{figure}}
\fi

In this section, we describe the program analysis component of \name's backend.
The {\em checker} performs liveness analysis~\cite{aho1986compilers},
and a \rev{lesser-known} program analysis technique
call\rev{ed} \rev{initialized variable} analysis\rev{, or definite assignment analysis}~\cite{moller2012static}.
These techniques are crucial for
efficiently identifying which cells are unsafe to execute due to stale references,
as well as
which cells help resolve staleness issues. We begin with background
before discussing the connection between these techniques
and staleness detection and resolution.

\subsection{Background}
\label{sec:analysis-bg}
{\em Liveness analysis}~\cite{aho1986compilers} is a program analysis technique for
determining whether the value of a variable at some point
is used later in the program. 
Although traditionally used by compilers to, for example,
determine how many registers need to be allocated at some point
during program execution, we use it to determine whether a cell
has references to stale symbols.
We \rev{also show (\S\ref{sec:highlights:nb}) how initialized variable analysis}
analysis~\cite{moller2012static}\rev{, a technique traditionally used by IDEs and linters
to detect potentially uninitialized variables, can be used}
to efficiently determine which cells to
\ifvldb
run in order to resolve staleness issues.
\else
to resolve staleness.
\fi

\begin{figure}[t]
\ifthesis
\begin{densecenter}
\begin{minipage}{0.5\textwidth}
\def\cellnumoffscfg{-4.0cm}
\else
\def\cellnumoffscfg{-4.3cm}
\fi
\begin{lstjupyter}[cellnum=,leftmargin=7pt,rightmargin=4.4cm,cellnumoffs=\cellnumoffscfg,cellnumfill=none]{cfg}
 if num % 3 == 0:
     !\HB!foobar =  True
     s = 'foobar' !\HE!
 elif num % 3 == 1:
     !\HB!foo =  True
     s = 'foo' !\HE!
 else:
     !\HB!s = 'bar'!\HE!
 !\HB!print(s, foobar)!\HE!
\end{lstjupyter}
\begin{tikzpicture}[remember picture, overlay,
    every node/.append style={%
      align=center,
      line width=1pt,
      minimum height=10pt,
        font=\scriptsize\ttfamily,
        fill=none
    }%
]
\tikzset{myarrow/.style={->, >=latex', shorten >=1pt, thick}}
    \node [draw, diamond, aspect=2] at ([shift={(2.75cm,0)}]cfg-cell-marker) (entry) {num};
    \node [draw=blue, rectangle, below = 3mm of entry] (foo) {foo};
    \node [draw=pinegreen, rectangle, left = 5mm of foo] (foobar) {foobar};
    \node [draw=purple, rectangle, right = 5mm of foo] (bar) {bar};
    \node [draw=cyan, rectangle, below = 10mm of entry] (print) {print};
	\draw[myarrow,->] (entry.west) to [out=-160,in=90] node[near start,fill=none,anchor=east]{\tiny\%3==0} (foobar.north);
	\draw[myarrow,->] (entry.south) to [out=-90,in=90] node[near start,fill=none,anchor=east]{\tiny\%3==1} (foo.north);
	\draw[myarrow,->] (entry.east) to [out=-20,in=90] node[near start,fill=none,anchor=west]{\tiny\%3==2} (bar.north);
	\draw[myarrow,->] (foobar.south) to [out=-90,in=-180] (print.west);
	\draw[myarrow,->] (foo.south) to [out=-90,in=90] (print.north);
	\draw[myarrow,->] (bar.south) to [out=-90,in=0] (print.east);
    \node [below left = 1.4cm and 3mm of foobar.west, align=left, anchor=west] {\textsf{\textbf{LIVE}}:\ \ \,num, foobar\\ \rev{\textbf{\inited}}: s};

    \connectzm{cfg-1}{pinegreen}
    \connectzm{cfg-2}{blue}
    \connectzm{cfg-3}{purple}
    \connectzm{cfg-4}{cyan}
\end{tikzpicture}
\thesisdel{\vspace{2em}}
\ifthesis
\vspace{2em}
\end{minipage}
\fi
\ifthesis
\else
\vspace{-5pt}
\fi
\caption{Example liveness and \rev{initialized variable} analysis.\smacke{Label CFG nodes with \bcircle{1} etc if time permits.}}
\label{fig:liveness-deadness-example}
\ifthesis
\end{densecenter}
\fi
\end{figure}
\topic{Example}
\ifthesis
In \Cref{fig:liveness-deadness-example}, symbols \verb!num! and \verb!foobar! are
live at the top of the cell, since the value for each at the top of the cell can be used in some path of the control flow graph (CFG).
In the case of \verb!num!, the (unmodified) value is used in the conditional.
In the case of \verb!foobar!, while one path of the CFG modifies it,
the other two paths leave it unchanged by the time it is used in the \verb!print! statement;
hence, it is also live at the top of the cell.
The symbol that is not live at cell start is \verb!foo!, since it is only ever assigned and never used,
and \verb!s!, since every path in the CFG assigns to \verb!s!.
We call symbols such as \verb!s! that are assigned in every path of the CFG {\em dead}
once they reach the end of the cell.
\else
In \Cref{fig:liveness-deadness-example}, symbols \verb!num! and \rverb|foobar| are
live at the top of the cell, since the value for each at the top of the cell can be used in some path of the control flow graph (CFG).
In the case of \verb!num!, the (unmodified) value is used in the conditional.
In the case of \rverb!foobar!, while one path of the CFG modifies \rev{it},
\rev{the other} two paths \rev{leave it unchanged} by the time \rev{it is} used in the \verb!print! statement;
hence, \rev{it is} also live at the top of the cell.
The symbol that \rev{is} not live at cell start \rev{is} \verb!foo!, since it is only ever assigned and never used,
and \verb!s!, since every path in the CFG assigns to \verb!s!.
We call symbols such as \verb!s! that are assigned in every \rev{path} of the CFG {\em dead}
once they reach the end of the cell.
\fi


\subsection{Cell Oriented \rev{Analysis}}
\label{sec:cell-liveness:nb}

We now describe how we relate liveness, which is traditionally applied in
the context of a single program, to a notebook environment. 
\techreport{In brief,
we treat each cell as if it is an individual program when performing
various program analyses. 
We formalize these notions below.}

\begin{definition}[Live symbols]
\label{def:livesym:nb}
Given a cell $c$ and some symbol \verb!s!, we say that \verb!s! is
{\em live in $c$} if
there exists some execution path in $c$ in which the value of \verb!s! at
the start of $c$'s execution is used later in $c$.
\end{definition}
In other words, \verb!s! is live in $c$ if,
treating $c$ as a standalone program, \verb!s!
is live in the traditional sense at the start of $c$.
We already saw in \Cref{fig:liveness-deadness-example} that
the live symbols in the example cell are \verb!num!, \verb!fiz!, and \verb!buz!.
For a given cell $c$, we use $\live(c)$ to denote the
\ifvldb
set of all
\fi
live symbols in $c$.

We are also interested in {\em dead symbols} that are
(re)defined in every branch by the time execution reaches the end
of a given cell $c$.
\begin{definition}[Dead symbols]
\label{def:deadsym:nb}
Given a cell $c$ and some symbol \verb!s!, we say that \verb!s! is
{\em dead in $c$} if, by the time control reaches the
end of $c$, every possible path of execution in $c$ overwrites \verb!s! in
a manner independent of the current value of \verb!s!.
\end{definition}
Denoting such symbols as $\dead(c)$, we will see in
\Cref{sec:highlights:nb} the role they play in assisting in the resolution of staleness issues.

\topic{Staleness and Freshness of Live Symbols in Cells}
Recall that symbols are augmented with additional lineage and timestamp metadata
computed by the tracer (\S\ref{sec:lineage:nb}). We can thus additionally refer to
the set $\stale(c) \subseteq \live(c)$, the set of stale symbols
that are live in $c$. When this set is nonempty, we say that cell $c$ itself is stale:

\begin{definition}[Stale cells]
\label{def:stalecell:nb}
A cell $c$ is called {\em stale} if
there exists some
$\verb!s! \in \live(c)$ such that \verb!s! is
stale; \ie, $c$ has a live reference to some stale symbol.
\end{definition}
\noindent
A major contribution of \name is to identify cells that are stale
and preemptively warn the user about them.

Note that a symbol can be stale regardless of whether it is live
in some cell. Given a particular cell $c$, we can also categorize
symbols according to their lineage and timestamp metadata as they
relate to $c$. For example, when a non-stale symbol \verb!s! that is live
in $c$ is more ``up-to-date'' than $c$, then we say that it is {\em fresh} with respect to $c$:

\begin{definition}[Fresh symbols]
\label{def:freshsym:nb}
Given a cell $c$ and some symbol \verb!s!, we say that \verb!s! is 
{\em fresh \wrt $c$} if (i) \verb!s! is not stale,
and (ii) $\ts{s} > \ts{$c$}$.
\end{definition}

We can extend the notion of fresh symbols to cells just as we did for
stale symbols and stale cells:

\begin{definition}[Fresh cells]
\label{def:freshcell:nb}
A cell $c$ is called {\em fresh} if it (i) it is not stale, and (ii) it
contains a live reference to one or more fresh symbols; that is,
$\exists \verb!s! \in \live(c)$ such that \verb!s! is fresh
with respect to $c$.
\end{definition}

\topic{Example}
Consider a notebook with three cells run in sequence, with
code \verb!a=4!, \verb!b=a!, and \verb!c=a+b!,
respectively, and suppose
the first cell is updated to be \verb!a=5! and rerun. The third cell
contains references to \verb!a! and \verb!b!, and although \verb!a! is
fresh, \verb!b! is stale, so the third cell is
not fresh, but stale. On the other hand, the second cell
contains a live reference to \verb!a! but no live references
to \verb!b!, and is thus fresh.

\vspace{3.5pt}
As we see in our experiments (\S\ref{sec:experiments:nb}), fresh cells
are oftentimes cells that users wish to re-execute; another major
contribution of \name is therefore
to automatically identify such cells.
In fact, in the above example, rerunning the second cell
resolves the staleness issue present in the first cell.
That said, running any other cell that assigns to \verb!b!
would also resolve the staleness issue, so staleness-resolving
cells need not necessarily be fresh. Instead, fresh cells can
be thought of as resolving staleness in cell output,
as opposed to resolving staleness in some symbol.
We study such staleness-resolving cells next.

\topic{Cells that Resolve Staleness}
We have already seen how liveness checking can help users to identify
stale cells. Ideally, we should also identify cells whose execution
would ``freshen'' the stale variables that are live in some cell $c$,
thereby allowing $c$ to be executed without potential errors due to staleness.
We thus define {\em refresher cells} as follows:

\begin{definition}[Refresher cells]
\label{def:refreshercell:nb}
A non-stale cell $c_r$ is called {\em refresher}
if there exists some other stale cell $c_s$ such that
\papertext{\vspace{-0.15em}}
\[ \stale(c_s) - \stale(c_r \oplus c_s) \ne \emptyset \]%
\papertext{\vspace{-0.15em}}%
where $c_r \oplus c_s$ denotes the concatenation of cells $c_r$ and $c_s$.
That is, the result of merging $c_r$ and $c_s$ together
has fewer live stale symbol references than does $c_s$ alone.
\end{definition}
\aditya{This is a bit funky. Does this mean
$c_r$ and $c_s$ are executed at the same timestamp?
Having trouble resolving this \wrt \Cref{def:stalesym:nb}.}

\noindent Intuitively, if we were to submit a refresher cell for execution,
we would reduce the number of stale symbols live in some other cell (possibly to 0).
Note that a refresher cell may or may not be fresh.

In addition to identifying stale and fresh cells, a final major contribution
of \name is the {\em efficient} identification of refresher cells.
We will see in \Cref{sec:highlights:nb} that scalable computation of such cells
requires \rev{initialized} analysis \rev{to} compute dead symbols\papertext{.}\techreport{,}
\iftech
which we describe \techreport{in detail} next.
\fi

\iftech
\else
\topic{\rev{Initialized Variable Analysis}}
\rev{Initialized variable analysis~\cite{moller2012static} can be thought of as the ``inverse''
of liveness analysis. While liveness analysis is a ``backwards-may'' technique for
computing symbols whose non-overwritten values ``may'' be used in a cell, initialized
analysis is a ``forwards-must'' technique that computes symbols that will be
``definitely assigned'' by the time control reaches the end of a cell.
Please see the technical report~\cite{nbtech} for a detailed discussion of initialized
analysis; we leverage it within the context of \name
to determine whether a non-stale cell will overwrite any stale symbols,
which turns out to be an efficient mechanism for computing refresher cells (\S\ref{sec:efficient-refresher}).}
\fi

\iftech
\subsection{\rev{Initialized Variable Analysis}}
\aditya{This is effectively unreadable. Give intuition and toss details and algos to appendix.}
\smacke{I would rather not do this; it looks rather evasive imho. I spent some effort improving intuition;
I do think the side-by-side comparison is helpful. Liveness analysis is fairly standard/textbook so I don't think
this is asking too much of reviewers.}
\label{sec:inverse-liveness}
Recall that we use liveness analysis to find ``live'' symbols whose values
at the start of each cell contribute to the computation performed in the
cell, and we use inverse liveness to find ``dead'' symbols whose values
at the end of the each will have definitely been overwritten by the
time control reaches the end of the cell.
A working knowledge of traditional liveness analysis
is a prerequisite for for understanding our inverse liveness technique;
we refer the reader to, \eg, Aho \etal~\cite{aho1986compilers}
for any review necessary.

\topic{Dataflow Equations}
Inverse liveness is a fixed-point method for solving the following set of
dataflow equations:
\papertext{\vspace{-.15em}}
\begin{align*}
    \dead_{out}[s] \quad&=\quad \big(\killed[s]-\use[s]\big) \quad\cup\quad \dead_{in}[s] \\
    \dead_{in}[s]  \quad&=\quad \bigcap_{s'\in\textsf{predecessors of $s$}}\dead_{out}[s']
\end{align*}
\papertext{\vspace{-.25em}}
That is, a symbol is dead in statement $s$ (i) if it is defined as a function,
(ii) if it appears on the \papertext{\lhs}\techreport{left hand side} of an assignment (but not the \papertext{\rhs)}\techreport{right hand side)},
or (iii) if it is dead in all predecessor statements in the control flow graph.

\topic{Intuition}
Traditional liveness analysis initializes each statement
at the minimum point of a lattice (\ie, the empty set).
Each statement's set of used symbols ($\use[s]$)
are then iteratively propagated
in the reverse direction of control until a fixed point is reached\papertext{.}\techreport{,}
\iftech
thereby solving the following set of dataflow equations:
\begin{align*}
  \live_{in}[s] = \use[s] \cup (\live_{out}[s] - \killed[s]) \\
  \live_{out}[s] = \bigcup_{s'\in\textsf{successors of $s$}}\live_{in}[s']
\end{align*}
\fi

If the live variables propagated during liveness checking can be thought
of as electrons, then the dead variables propagated during inverse liveness
can be thought of as ``holes'', using a metaphor from electronics.
Nearly every decision made over the course of inverse liveness analysis
is the ``inverse'' of decisions made during liveness analysis. For example,
in inverse liveness, each statement's set of dead symbols is initialized to {\em everything}
(\ie, at lattice maximum),
and inverse liveness uses set intersection ($\cap$) as the lattice meet operator
instead of set union ($\cup$) when propagating dead variables between statements,
since symbols can be dead at cell bottom only if they are overwritten in {\em every}
branch of control. \papertext{The salient differences between liveness
and inverse liveness are highlighted in \Cref{alg:liveness:nb,alg:deadness:nb};
please see the technical report~\cite{nbtech} for more details.}

\topic{Comparing Liveness and Inverse Liveness}
A pseudocode description of our inverse liveness
checker is given in \Cref{alg:deadness:nb}.
We also give a description of a textbook liveness checker
to the left in \Cref{alg:liveness:nb} for contrast.
In particular, we note that the two algorithms are nearly
identical excepting a few key differences:
\begin{denselist}
\item On {\color{linkpurple}line \ref{line:dead-init}},
      $\dead_{out}$ is initialized to {\em every}
      symbol, instead of $\emptyset$;
\item On {\color{linkpurple}line \ref{line:dead-incoming}}
      of \Cref{alg:deadness:nb},
      we take the intersection of outgoing dead
      symbols in predecessor nodes, instead of the union of incoming
      live symbols in successor nodes as in \Cref{alg:liveness:nb};
\item On {\color{linkpurple}line \ref{line:dead-outgoing}}
      of \Cref{alg:deadness:nb}, we compute outgoing dead symbols as the
      union of symbols killed in the current node with incoming dead symbols,
      whereas \Cref{alg:liveness:nb} computes incoming live symbols as the
      union of symbols used in the current node with non-killed
      outgoing live symbols;
\item The dataflows is from top to bottom in \Cref{alg:deadness:nb},
      while it is from bottom to top in \Cref{alg:liveness:nb}.
\end{denselist}
These differences underscore the role \Cref{alg:deadness:nb}
plays as an inversion of liveness analysis.
\fi

\subsection{\rev{Resolving Live Symbols}}
\label{sec:resolution:nb}

\rev{In many cases, it is possible to determine the set of live symbols
in a cell with high precision purely via static analysis. In some cases,
however, it is difficult to do so without awareness of additional runtime
data. To illustrate, consider the example below:}

\begin{figure}[H]
\ifvldb
\else
\vspace{-1em}
\fi
\begin{lstjupyter}[cellnum={[1]}]{res1}
  x = 0
  def f(y):
      return x + y
  lst = [f, lambda t: t + 1]
\end{lstjupyter}

\begin{lstjupyter}[cellnum={[2]}]{res2}
  print(lst[1](2))
\end{lstjupyter}
\end{figure}

\rev{Whether or not symbol} \rverb!x! \rev{should be considered live at the top of the second cell
depends on whether the call to} \rverb!lst[1](2)! \rev{refers to
the list entry containing the lambda, or the entry containing function}
\rverb!f!\rev{. In this case, a static analyzer might be able to infer that}
\rverb!lst[1]! \rev{does not reference} \rverb!f! \rev{and that} \rverb!x!
\rev{should therefore not be considered live
at the top of cell 2 (since there is no call to function} \rverb!f!\rev{, in whose
body} \rverb!x! \rev{is live), but doing so in general is challenging due to Rice's theorem. Instead of
doing so purely statically, \name performs an extra resolution step, since it
can actually examine the runtime value of} \rverb!lst[1]! \rev{in memory. This allows
\name to be more precise about live symbols than a conservative approach would be,
which would be forced to consider} \rverb!x! \rev{as live even though} \rverb!f!
\rev{is not referenced by} \rverb!lst[1]!\rev{.}

\section{Cell Highlights}
\label{sec:highlights:nb}

In this section, we describe how to combine the lineage metadata
\techreport{from \Cref{sec:lineage:nb}} with the output of \techreport{\name's }\papertext{the }static checker
to highlight cells of interest.

\subsection{Highlight Abstraction}
\label{sec:highlight-abstraction}
We begin by defining the notion of
{\em cell highlights} in the abstract before discussing concrete examples,
how they are presented, and how they are computed.

\begin{definition}[Cell highlights]
\label{def:highlights:nb}
Given a notebook $N$ abstractly defined as an ordered set of
cells $\{c_i\}$, a set of {\em cell highlights} $\hl$ is a subset
of $N$ comprised of cells that are semantically related in some way
at a particular point in time.
\end{definition}
More concretely, we will consider the following cell highlights:
\begin{denselist}
\item $\hls$, the set of stale cells in a notebook;
\item $\hlf$, the set of fresh cells in a notebook; and
\item $\hlr$, the set of refresher cells in a notebook.
\end{denselist}
Note that these sets of cell highlights are all implicitly
indexed by their containing notebook's execution counter.
When not clear from context we write $\hlst$, $\hlft$, and $\hlrt$ (respectively)
to make the time dependency explicit.
Along these lines, we are also interested in the following
``delta'' cell highlights:
\begin{denselist}
\item $\hlfpt = \hlft - \hlft[t-1]$ (new fresh cells); and
\item $\hlrpt = \hlrt - \hlrt[t-1]$ (new refresher cells)
\vspace{2pt}
\end{denselist}
again omitting superscripts when clear from context.

\topic{Interface}
We have already seen from the example in \Cref{fig:highlight-example}
that stale cells are given \unsafehl{staleness warnings} to the left of
the cell, and refresher cells are given \refresherhl{cleanup suggestions}
to the left of the cell.
\techreport{The current version of }
\name \techreport{as of this writing (0.0.49)}
also augments fresh cells with \refresherhl{cleanup suggestions} of the same
color as that used for refresher cells. Overall, the fresh and refresher highlights
are intended to steer users toward cells that they may wish to re-execute, and
the stale highlights are intended to steer users away from cells that they may wish to avoid,
intuitions that we validate in our empirical study (\S\ref{sec:experiments:nb}). 
\techreport{Experimenting
with presentation techniques for the various sets $\hlstar$ 
is an interesting avenue that we leave to future work.}

\andrew{This might not be the right focus for this paper, though it might be worth mentioning non-trivial design insights that went into the current interface design. I imagine there are several designs that were thrown out as unhelpful, and that this one was seen as a nice combination of discoverable, understandable, actionable, and unobtrusive. Is there any intellectual work that went into the design that would save time for future tool builders of similar tools? If so, it might be worth mentioning here, and alluding to in the introduction.}\dlee{I think this would be an interesting discussion to add if space permits.}

\topic{Computation}
Computing $\hls$ and $\hlf$ is straightforward: for each cell $c$,
we simply run a liveness checker \techreport{(\Cref{alg:liveness:nb})} to determine
$\live(c)$, and then perform a metadata lookup for each symbol $\verb!s!\in\live(c)$
to determine whether \verb!s! is fresh \wrt $c$ or stale.
\iftech
The manner in which \name computes refresher cells
\else
\rev{Refresher cell computation}
\fi
deserves a more thorough treatment that we consider next.

\subsection{Computing Refresher Cells Efficiently}
\label{sec:efficient-refresher}

Before we discuss how \name uses \rev{\techsub{an}{the} initialized variable} checker
from \Cref{sec:analysis:nb} to efficiently compute refresher cells,
consider how one might design an algorithm to compute refresher cells directly from
\Cref{def:refreshercell:nb}. The straightforward way is to loop over all non-stale
cells $c_r \in N - \hls$ and compare whether $\stale(c_r\oplus c_s)$ is smaller than $\stale(c_s)$.
In the case that $\hls$ and $N-\hls$ are similar in size, this requires performing
$\bigo{|N|^2}$ liveness analyses, which would create unacceptable latency in the case of large notebooks.
\techreport{This inefficient approach is depicted in \Cref{alg:refresher-naive}.}

\iftech
\begin{algorithm}[t]
\caption{Computing refresher cells \naively}\label{alg:refresher-naive}
\algrenewcomment[1]{\texttt{/*} #1 \texttt{*/} \\}
\SetKwInOut{Input}{Input}
\SetKwInOut{Output}{Output}
\SetKwRepeat{Do}{do}{while}
\SetKwRepeat{Repeat}{repeat}{until}
\SetKwProg{myfunc}{function}{}{}
{\scriptsize
\Input{Notebook $N$, stale cells $\hls \subseteq N$}
\Output{Refresher cells $\hlr$}
$\hlr \gets \emptyset$\;
\ForEach{$c_s\in\hls$}{
  $\stale(c_s) \gets$ live, stale symbols in $c_s$\;
  \ForEach{$c_r\in N-\hls$}{
    $\stale(c_r\oplus c_s) \gets$ live, stale symbols in $c_r\oplus c_s$\;
    \If{$\stale(c_s) - \stale(c_r\oplus c_s) \ne \emptyset$}{
      $\hlr \gets \hlr \cup \{c_r\}$\;
    }
  }
}
\Return{$\hlr$}\;
}
\end{algorithm}

\begin{algorithm}[t]
\caption{Computing refresher cells efficiently}\label{alg:refresher:nb}
\algrenewcomment[1]{\texttt{/*} #1 \texttt{*/} \\}
\SetKwInOut{Input}{Input}
\SetKwInOut{Output}{Output}
\SetKwRepeat{Do}{do}{while}
\SetKwRepeat{Repeat}{repeat}{until}
\SetKwProg{myfunc}{function}{}{}
{\scriptsize
\Input{Notebook $N$, stale cells $\hls \subseteq N$,\\
stale and live symbols $\stale(c_s), \forall c_s\in\hls$,\\
dead symbols $\dead(c_r), \forall c_r\in N-\hls$}
\Output{Refresher cells $\hlr\subseteq N$}
$\invdead[s] \gets \emptyset, \forall s\in\dead(c_r), \forall c_r\in N - \hls$\;
\ForEach{$c_r\in N - \hls$}{%
	\ForEach{$\code{s}\in\dead(c_r)$}{%
		$\invdead[s] \gets \invdead[s] \cup \{c_r\}$\;
	}
}
$\hlr \gets \emptyset$\;
\ForEach{$c_s\in\hls$}{
  \ForEach{$s\in\stale(c_s)$}{
    $\hlr \gets \hlr \cup \invdead[s]$\;
  }
}
\Return{$\hlr$}\;
}
\end{algorithm}

\fi

By leveraging \rev{an initialized variable} checker, it turns out that
we can check whether $\stale(c_s)$ and $\dead(c_r)$ have any overlap
instead of performing liveness analysis over $c_r\oplus c_s$ and checking
whether $\stale(c_r\oplus c_s)$ shrinks.
We state this formally as follows:

\begin{theorem}
\label{thm:dead-to-refresher}
Let $N$ be a notebook, and let $c_s \in \hls \subseteq N$.
For any other $c_r\in N-\hls$, the following equality holds:
\[ \stale(c_s) - \stale(c_r\oplus c_s) = \dead(c_r) \cap \stale(c_s) \]
\end{theorem}
\iftech
\begin{proof}
We show each side of the equality is a subset of the other side.
First, suppose some stale symbol \verb!x! is live in $c_s$ but not in $c_r\oplus c_s$.
Then, at the point where control transfers from $c_r$
to $c_s$ in the outermost scope, every path of execution will definitely have
redefined \verb!x!.\footnote{Note that there is no way for control to transfer from $c_r$
to $c_s$ in the outermost scope except at the point where $c_r$ and $c_s$ meet lexically
(technically, $c_r$ could call a function defined
in $c_s$, but if this were to occur, control would not be at the outermost scope).}
Otherwise, there would exist a path in $c_r$ wherein \verb!x! is not redefined,
and because \verb!x! is live at the top of $c_s$, it would also be live at the top of $c_r\oplus c_s$.
As such, $\verb!x! \in \dead(c_r)$ by definition. Furthermore, $\verb!x!\in\stale(c_s)$ by our initial assumption,
so $\verb!x!\in\dead(c_r)\cap\stale(c_s)$.

Conversely, suppose some stale symbol \verb!x! is live in $c_s$ but dead in $c_r$.
By definition, every path of execution in $c_r$ redefines \verb!x!. We would like
to say that \verb!x! is not live in $c_r\oplus c_s$, but deadness in $c_r$ does not
preclude liveness in $c_r$ (if, \eg, \verb!x! is used in some path of $c_r$ before
it is redefined later). Thus, it is only true that \verb!x! is not live if $c_r\oplus c_s$
if it is also not live in $c_r$. In fact, \verb!x! is not live in $c_r$ because $c_r\notin\hls$; \ie, $c_r$ has no
live stale symbols by assumption, and \verb!x! is stale; thus \verb!x! is both live in $c_s$
and not live in $c_r\oplus c_s$; \ie, $\verb!x!\in\stale(c_s) - \stale(c_r\oplus c_s)$,
which is what we needed to show to complete the proof.
\end{proof}
\else
Please see the technical report~\cite{nbtech} for a proof. $\square$
\fi

\noindent \Cref{thm:dead-to-refresher} relies crucially on the fact that
the CFG of the concatenation of two cells $c_r$ and $c_s$ into $c_r\oplus c_s$
will have a ``choke point'' at the position where control transfers from $c_r$ into $c_s$,
so that any symbols in $\dead(c_r)$ cannot be ``revived'' in $c_r\oplus c_s$.

\topic{Computing $\hlr$ Efficiently}
Contrasted with taking $\bigo{|N|^2}$ pairs $c_s\in\hls$, $c_r\in N-\hls$ and checking
liveness on each concatenation $c_r\oplus c_s$, \Cref{thm:dead-to-refresher}
instead allows us compute the set $\hlr$ as
\begin{equation}
\label{eqn:refresher-formula}
\ifthesis
\bigcup_{c_s\in\hls}\bigcup_{\texttt{s}\in\stale(c_s)}\big\{c_r\in N-\hls: \texttt{s}\in\dead(c_r)\big\}
\else
\iflib
\bigcup_{c_s\in\hls}\bigcup_{\verb!s!\in\stale(c_s)}\big\{c_r\in N-\hls: \verb!s!\in\dead(c_r)\big\}
\else
\bigcup_{c_s\in\hls}\bigcup_{\texttt{s}\in\stale(c_s)}\big\{c_r\in N-\hls: \texttt{s}\in\dead(c_r)\big\}
\fi
\fi
\end{equation}
{\color{linkpurple}Equation}~\ref{eqn:refresher-formula}
can be computed efficiently
\iftech
according to \Cref{alg:refresher:nb}, which creates an
\else
by creating
\fi
inverted index that maps dead symbols to their containing cells ($\invdead$)
in order to efficiently compute the inner set union.
Furthermore,
\iftech
\Cref{alg:refresher:nb}
\else
this approach
\fi
only requires $\bigo{|N|}$ liveness analyses and $\bigo{|N|}$ \rev{initialized variable} analyses as preprocessing\rev{,
translating to significant latency reductions in our benchmarks (\S\ref{sec:benchmarks:nb}).}

\section{Empirical Study}
\label{sec:experiments:nb}

We now  evaluate
\name's ability to highlight unsafe cells, as well as
cells that resolve safety issues (refresher cells).
We do so by replaying \numworking real notebook
sessions and measuring how the cells highlighted by \name
correlate with real user actions. After describing data collection (\S\ref{sec:data:nb})
and our evaluation metrics (\S\ref{sec:metrics:nb}), we present our
\rev{quantitative} results \rev{(\S\ref{sec:results:nb} and \S\ref{sec:benchmarks:nb})\papertext{.}\techreport{,}
\iftech
followed by a qualitative comparison with other systems
informed by real examples from our data (\S\ref{sec:comparison:nb} and \S\ref{sec:case:nb}).}
\fi

\subsection{Notebook Session Replay Data}
\label{sec:data:nb}
We now describe our data collection and session replay efforts.

\topic{Data Scraping}
The \verb!.ipynb! json format contains a static snapshot of the code present
in a computational notebook and lacks explicit interaction data, such as how
the code present in a cell evolves, which cells are re-executed, and the order
in which cells were executed.\papertext{ }\techreport{\footnote{The cell counter in a\spacecodespace{.ipynb}file
only contains the latest executed cell version for each cell, and says nothing
about how executions of earlier iterations of the cell are ordered \wrt others.}}
Fortunately, Jupyter's IPython kernel
implements a history mechanism that includes information about
individual cell executions in each session, including the source code and execution counter for every cell execution.
We thus scraped \verb!history.sqlite! files from \numrepos repositories
files using GitHub's API~\cite{ghapi}, from which we successfully extracted
\numhistories such files.
In total, these history files contained execution logs
for $\approx$ \numsessions notebook sessions, out of which we were
able to collect metrics for \numworking
after conducting the filter and repair steps described next.

\topic{Notebook Session Repair}
\rev{Many of the notebook sessions were impossible to replay
with perfect replication of the session's original behavior (due to, \eg, missing files).}
To cope, we adapted ideas from Yan \etal~\cite{yan2020auto}
to repair sessions wherever possible.
\iftech
Specifically, we took the following measures:
\begin{denselist}
\item Since \name runs on Python 3, we used the \verb!2to3! tool~\cite{2to3} whenever we encountered Python 2 code.
\item To deal with differing APIs used by different versions of the same library (\eg, \verb!scikit-learn!),
      we first gathered all the import statements for each library and tried to execute them under different versions
      of the aforementioned library, using the version that minimized import errors to finally replay the session.
\item We normalized all path-like strings to point to the same directory, to prevent invalid accesses to nonexistent directories.
\item We used the Kaggle API to search for and attempt to download any \verb!csv! files referenced by each session.
\item We removed any lines or cells that attempted to run system commands through Jupyter's line (\resp cell) magic functionality.
\item We executed the line magic \verb!%matplotlib! \verb!inline! before replaying a session to avoid rendering matplotlib charts
      with \verb!Qt!.
\end{denselist}
\else
Please see our technical report~\cite{nbtech}
for details for repair and filtering (below).
\fi

\topic{Session Filtering}
Despite these efforts, we were unable to reconstruct some sessions to their original fidelity due to various environment discrepancies.
Furthermore, certain sessions
had few
cell executions and appeared to be random tinkering.
\iftech
We therefore
filtered out sessions fitting any of the following criteria:
\begin{denselist}
\item Sessions with fewer than 50 cell executions;
\item Sessions that attempted to run shell commands;
\item Sessions that solicited user input via \verb!readline! or other means;
\item Sessions that attempted to connect to external services (\eg AWS, Spark, Postgres, MySQL, \etc);
\item Sessions that attempted to read nonexistent files
      (or those that could not be found using the Kaggle API).
\end{denselist}
After these steps, we were left with \numrepaired replayable sessions.
\else
We therefore filtered out undesirable sessions,
 after which we were left with
\numrepaired replayable sessions.
\fi
However, we were unable to gather meaningful metrics
on more than half of the sessions we replayed because of exceptions thrown upon many cell executions.
We filtered these in post-processing by removing data for any session with more
than \excthresh of cell executions resulting in exceptions.
\dlee{It is unclear why we chose to throw out these sessions completely, instead of truncating starting at the point when exceptions are thrown. Since notebooks have sequential dependencies, typically if something throws an exception, many cells downstream also breaks.}
\smacke{Probably not going to resolve this comment because
the real reason I didn't do the fancier thing is laziness :)}

After the repair and filtration steps, we extracted metrics from a total of \numworking sessions.
Our
\ifvldb
repair, filtering, and replay
\fi
scripts are available on GitHub~\cite{nbreplay}.
\iftech

\topic{Environment}
All experiments were conducted on a 2019 MacBook Pro with 32 GiB RAM and a Core i9 processor
running macOS 10.14.5, Python 3.7, and \name 0.0.49. We replayed notebook sessions
in a container instance to ensure our local files would not be compromised in the
event of intentionally or unintentionally malicious code present in the sessions we scraped.
\else
\fi

\subsection{Metrics}
\label{sec:metrics:nb}
\rev{Besides conducting benchmark experiments to measure overhead associated
with \name{} (\S\ref{sec:benchmarks:nb}), the primary goal of our empirical study is
to evaluate our system and interface design choices from the previous sections by
testing two hypotheses.}
Our first hypothesis (i)
is that \rev{{\em cells with staleness issues highlighted by \name are likely to be
avoided by real users}, suggesting that these cells are indeed unsafe to execute.}
Our second hypothesis (ii)
is that {\em \rev{fresh and refresher cells highlighted by \name} are more likely to be
selected for re-execution}, \rev{indicating that these suggestions can help reduce}
cognitive overhead for users
trying to choose which cells to re-execute.
To \rev{test these hypotheses}, we introduce the notion of
{\em \metric} for cell highlights.
\begin{definition}[Predictive Power]
Given a notebook $N$ with a total of $|N|$ cells,
\aditya{<- that could be executed next? what about new cells?}
the id of the next cell executed $c$,
and a non-empty set of cell highlights $\hl$ (chosen before $c$ is known),
the {\em predictive power} of $\hl$ is defined as $\pp(\hl) = \indic{c\in\hl}\cdot |N|/|\hl|$.
\end{definition}
\noindent Averaged over many measurements,
predictive power \rev{assesses} how many more times more likely
a cell from some set of highlights $\hl$ is to be picked for re-execution,
compared to random cells.

\topic{Intuition}
To understand predictive power, consider a set of highlights $\hl$
chosen uniformly randomly without replacement from the entire set of available cells.
In this case,
\iftech
\[ \Earg{\indic{c\in\hl}} = \Parg{c\in\hl} = |\hl| / |N| \]
\else
$\Earg{\indic{c\in\hl}} = \Parg{c\in\hl} = |\hl| / |N|$,
\fi
so that the predictive power of $\hl$ is $(|\hl|/|N|)\cdot(|N|/|\hl|)=1$.
This holds for any number of cells in the set of highlights $\hl$, even
when $|\hl|=|N|$.
Increasing the size of $\hl$
increases the chance for a nonzero predictive power, but it also decreases
the ``payout'' when $c \in \hl$.
For a fixed notebook $N$, the maximum possible predictive power for $\hl$ occurs when
$\hl = \{c\}$, in which case $\pp(\hl) = |N|$.
\aditya{This metric is not normalized...}

\topic{Rationale} Our goal in introducing predictive power
is not to give a metric that we then attempt to optimize; rather,
we merely want to see how different sets of cell highlights correlate
with real user behavior. In some sense, any $\pp(\hl) \ne 1$ is interesting:
$\pp(\hl) < 1$ indicates that users tend to avoid $\hl$, and $\pp(\hl) > 1$
indicates that users tend to prefer $\hl$. For the different sets of
cell highlights $\{\hlstar\}$ introduced in \Cref{sec:highlights:nb}, each $\pp(\hlstar)$
helps us to make this determination.

\topic{Gathering measurements}
The session interaction data available in the scraped history files
only contains the submitted cell contents for each cell execution,
and unfortunately lacks cell identifiers. Therefore, we attempted
to infer the cell identifier as follows: for each cell execution,
if the cell contents were $\geq\leventhresh$ similar to a previously
submitted cell (by Levenshtein similarity),
we assigned the identifier of that cell; otherwise, we assigned a new identifier.
Whenever we inferred that an existing cell was potentially edited and re-executed,
we measured predictive power for various highlights $\hlstar$ when such highlights
were non-empty.
Across the various highlights, we computed the average of such predictive powers
for each sessions, and the averaged the average predictive powers across all
sessions, reporting the result as $\AVG(\pp(\hlstar))$ for each $\hlstar$ (\S\ref{sec:results:nb}).

\topic{Highlights of Interest}
We gathered metrics for $\hls$, $\hlf$, $\hlfp$, $\hlr$, and $\hlrp$, which
we described earlier in \Cref{sec:highlights:nb}. Additionally, we also gathered metrics
for the following ``baseline highlights'':
\begin{denselist}
\item $\hln$, or the {\em next cell highlight}, which contains only the $k+1$ cell (when applicable) if cell $k$ was the previous cell executed; and \aditya{<- Do cells have order? You just said they didn't.}
\item $\hlrr$, or the {\em random cell highlight}, which simply picks a random cell from the list of existing cells.
\end{denselist}
We take measurements for $\hln$ because picking the next cell in a notebook is a common choice,
and it is interesting to see how its predictive power compares with cells highlighted by the
\name frontend such as $\hlf$ and $\hlr$. We also take measurements for $\hlrr$ to validate
via Monte Carlo simulation the claim that random cells $\hlrr$ should satisfy $\pp(\hlrr) = 1$ in expectation.


\subsection{\rev{Predictive Power} Results}
\label{sec:results:nb}

\iftech
\begin{figure}[t]
\begin{densecenter}
\includegraphics[width=\thesissub{\linewidth}{0.7\linewidth}]{\figs/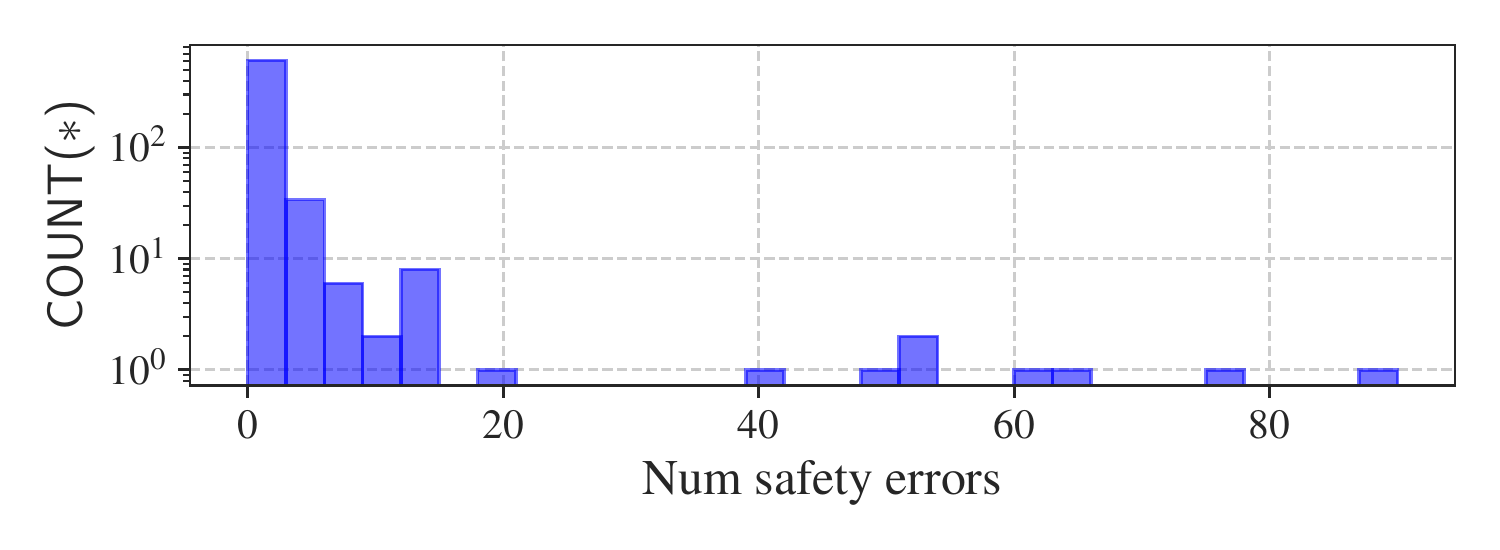}
\thesisdel{\vspace{-2em}}
\caption{Histogram showing distribution of safety errors across sessions.}
\label{fig:safety-error-hist}
\end{densecenter}
\end{figure}
\fi

\begin{figure}[t]
\begin{densecenter}
\ifvldb
\else
\thesisdel{\vspace{-1em}}
\fi
\resizebox{\thesissub{\linewidth}{0.7\linewidth}}{!}{\input{\figs/compare-pp-with-without-safety-issues.pgf}}
\thesisdel{\vspace{-2em}}
\ifvldb
\caption{Comparing \AVG($\pp(\hlstar)$) for sessions with/without safety issues.}
\else
\caption{\AVG($\pp(\hlstar)$) for sessions with/without safety issues.}
\fi
\label{fig:compare-pp-with-without-safety-issues}
\end{densecenter}
\end{figure}

\begin{table}[t]
\begin{densecenter}
\ifvldb
\iftech
\vspace{-1em}
\fi
\else
\fi
{\thesisdel{\scriptsize}\begin{tabular}{|c|c|c|c|c|c|c|c|}
\hline\rowcolor[HTML]{C5C5C5}{\bf Quantity} & $\hln$ & $\hlrr$ & $\hls$ & $\hlf$ & $\hlr$ & $\hlfp$ & $\hlrp$\\\hline
\AVG($\pp(\hlstar)$) &$2.64$ & $1.02$ & $\mathbf{0.30}$ & $2.83$ & $3.90$ & $\mathbf{9.17}$ & $6.20$\\\hline
\AVG($|\hlstar|$) & $1.00$ &  $1.00$ &  $2.71$ &  $3.73$ &  $2.31$ &  $1.73$ &  $1.81$\\\hline
\end{tabular}}
\caption{Summary of measurements taken for various highlight sets.}
\label{tab:hls-summary}
\end{densecenter}
\end{table}

In this section, we present the results of our empirical evaluation.
{\em Overall, \name discovered that \numunsafe sessions out of the \numworking
encountered staleness issues at some point},
underscoring the
\ifvldb
very real 
\else
\fi
need for a tool to prevent such errors.
Furthermore, we found that the ``positive'' highlights like $\hlf$ and $\hlr$
correlated strongly with user choices.

\topic{Predictive Power for Various Highlights}
We now discuss average $\pp(\hlstar)$ for the various $\hlstar$ we consider,
summarized in \Cref{tab:hls-summary}.

\frameme{{\bf \em Summary.}
\aditya{Couldn't follow this.}
Out of the highlights $\hlstar$ with $\pp(\hlstar) > 1$,
new fresh cells, $\hlfp$, had the highest predictive power,
while $\hln$ had the lowest (excepting $\hlrr$, which had $\pp(\hlrr) \approx 1$
as expected). $\hls$ had the lowest predictive power coming in at $\pp(\hls) \approx 0.30$,
suggesting that users do, in fact, avoid stale cells.}

\smacke{TODO: remove $\hlrr$ from table and briefly mention here instead.
Also add $\hlsp$ after rerunning experiments.}
We measured the average value of $\pp(\hls)$ at roughly $0.30$, which is the lowest
mean predictive power measured out of any highlights. One way to interpret this is that
users were more then $3\times$ {\em less} likely to re-execute stale cells than they
are to re-execute randomly selected highlights of the same size as $\hls$ ---
strongly supporting the hypothesis that users tend to avoid stale cells.

On the other hand, all of the highlights $\hln$, $\hlf$, $\hlr$, $\hlfp$, and $\hlrp$
satisfied $\pp(\hlstar) > 1$ on average, with $\pp(\hlfp)$ larger than the others
at $9.17$, suggesting that users are {\em more than $9\times$ more likely}
to select newly fresh cells to re-execute than they are to re-execute randomly selected
highlights of the same size as $\hlfp$. In fact, $\hln$ was the lowest non-random set
of highlights with mean predictive power $>1$, strongly supporting our design decision
of specifically guiding users to all the cells from $\hlf$ and $\hlr$ (and therefore to $\hlfp$ and $\hlrp$ as well)
with our aforementioned \refresherhl{visual cues}. Furthermore, we found that
no $|\hlstar|$ was larger than $4$ on average, suggesting that these cues
are useful, and not overwhelming.

Finally, given the larger predictive powers of $\hlfp$ and $\hlrp$, we plan to study interfaces
that present these highlights separately from $\hlf$ and $\hlr$ in future work.

\topic{Effect of Safety Issues on Predictive Power}
Of the \numworking sessions we replayed, we detected 1 or more safety issues (due to the user
executing a stale cell) in \numunsafe, while the majority (\numsafe) did not have safety issues.
\iftech
A histogram depicting the distribution of ``\# safety issues'' is given in \Cref{fig:safety-error-hist}.
\fi
We reveal interesting behavior by computing $\AVG(\pp(\hlstar))$ when restricted
to (a) sessions without safety errors, and (b) sessions with 1 or more safety errors,
depicted in \Cref{fig:compare-pp-with-without-safety-issues}.

\frameme{{\bf \em Summary.} For sessions with safety errors, users were more likely to
select the next cell ($\hln$), and less likely to select fresh or refresher cells
($\hlf$ and $\hlr$, respectively).}

\Cref{fig:compare-pp-with-without-safety-issues} plots $\AVG(\pp(\hlstar))$
for various highlight sets after faceting on sessions that did and did not
have safety errors. By definition, $\AVG(\pp(\hls)) = 0$ for sessions without
safety errors (otherwise, users would have attempted to execute one or more stale cells),
but even for sessions with safety errors, we still found $\pp(\hls)<1$ on average,
though not enough to rule out random chance.

Interestingly, we found that $\AVG(\pp(\hln))$ was significantly
higher for sessions with safety issues, suggesting that users were more likely
to execute the next cell without much thought.

Finally, we found that users were significantly {\em less} likely to choose
cells from $\hlf$, $\hlr$, or $\hlrp$ for sessions with safety errors. In fact,
users favored $\hln$ over $\hlr$ or $\hlf$ in this case.
Regardless of whether sessions had safety issues, however, $\hlfp$ and $\hlrp$
still had the highest mean predictive powers out of any of the highlights\papertext{.}\techreport{,}
\iftech
with $\AVG(\pp(\hlfp))$ relatively unaffected by safety issues.
\fi

\subsection{\rev{Benchmark Results}}
\label{sec:benchmarks:nb}

\iftech
\begin{figure}[t]
\begin{densecenter}
\resizebox{\thesissub{\linewidth}{0.7\linewidth}}{!}{\input{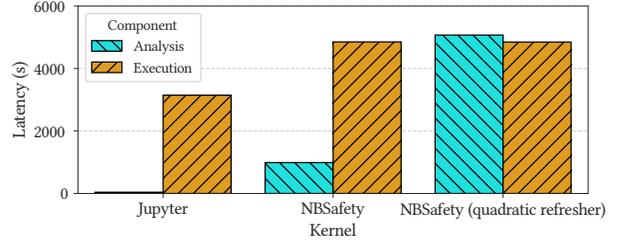}}
\thesisdel{\vspace{-2em}}
\caption{\rev{Comparing latencies of execution and \thesisdel{(if applicable)} analysis components for different kernels.}}
\label{fig:compare-component-latencies}
\end{densecenter}
\end{figure}
\fi

\begin{figure}[t]
\begin{densecenter}
\iftech
\ifthesis
\else
\vspace{-2.2em}
\fi
\fi
\ifvldb
\else
\vspace{-1em}
\fi
\resizebox{\thesissub{\linewidth}{0.7\linewidth}}{!}{\input{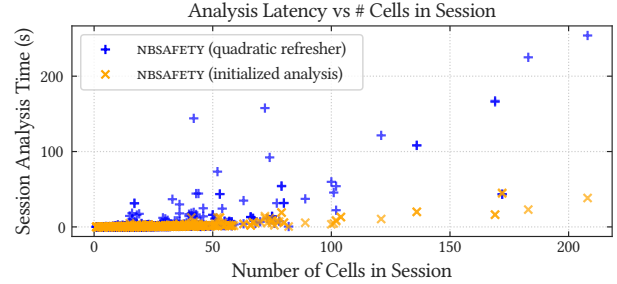}}
\thesisdel{\vspace{-2em}}
\caption{\rev{Measuring the impact of cell count on analysis latency for
\name with and without efficient refresher computation.}}
\label{fig:analysis-vs-cells}
\end{densecenter}
\end{figure}

\begin{table}[t]
\begin{densecenter}
\ifvldb
\iftech
\vspace{-1em}
\fi
\else
\ifthesis
\else
\vspace{-1.75em}
\fi
\fi
\rev{\thesisdel{\scriptsize}\begin{tabular}{|c|c|c|c|}
\hline\rowcolor[HTML]{C5C5C5}{\bf Approach} & Jupyter & \name & \name{} (quadratic refresher) \\\hline
\iftech
\else
Analysis Time (s) & $0$ & $990$ & $5070$ \\\hline
Execution Time (s) & $3150$ & $4850$ & $4850$ \\\hline
\fi
Total Time (s) & $3150$ & $5840$ & $9920$ \\\hline
Median Slowdown & $1\times$ & $1.44\times$ & $1.58\times$ \\\hline
\end{tabular}}
\caption{\rev{Summary of latency measurements.
Median slowdown measured on sessions that took
\ifvldb
longer than $5$ seconds
\else
$>5$ seconds
\fi
to execute in vanilla Jupyter.}}
\label{tab:latencies}
\end{densecenter}
\end{table}

\rev{Our benchmarks are designed to assess the additional overhead incurred
by our tracer and checker by measuring the end-to-end execution latency for
the aforementioned \numworking sessions, with and without \name.
Furthermore, we assess the impact of our initialized analysis
approach to computing refresher cells
\techreport{(\Cref{alg:refresher:nb})}
by comparing it with the \naive quadratic
\iftech
baseline (\Cref{alg:refresher-naive}).
\else
baseline (both discussed in \Cref{sec:highlights:nb}).
\fi%
}

\topic{\rev{Overall Execution Time}}
\rev{We summarize the time needed for various methods to replay the \numworking sessions
in our execution logs in \Cref{tab:latencies}, and furthermore faceted on the static analysis
and tracing / execution components in
\iftech
\Cref{fig:compare-component-latencies}.
\else
the same table.
\fi
We measured latencies for both vanilla Jupyter and \name, as well as for an ablation
that replaces the efficient refresher computation algorithm with the 
quadratic
\iftech
variant (\Cref{alg:refresher-naive}).
\else
variant.
\fi}

\frameme{\rev{{\bf \em Summary.}
The additional overhead introduced by \name is within the same order-of-magnitude as
vanilla Jupyter, taking less than $2\times$ longer to replay all \numworking sessions,
with typical slowdowns less than $1.5\times$. Without initialized analysis for refresher
computation, however, total reply time increased to more than $3\times$ the time taken
by \thesisdel{the vanilla} Jupyter\thesisdel{ kernel}.}}

\noindent \rev{Furthermore, we see from
\iftech
\Cref{fig:compare-component-latencies}
\else
\Cref{tab:latencies}
\fi
that refresher computation begins to dominate with the quadratic variant,
while it remains \thesissub{relatively minor}{small} for the linear variant based on initialized
analysis.}

\iftech
\rev{Although \name's tracer introduces some additional overhead compared to
the vanilla Jupyter kernel, we note that this overhead is relatively minor (less than $1.5\times$),
and that a less-optimized tracing implementation would have performed far worse. For example,
suppose} \rverb!lst! \rev{contains one million elements, and we materialize the output of a}
\rverb!map! \rev{operation, \eg} \rverb!lst = list(lst.map(f))!\rev{. Using Python's tracing mechanism
directly, this would produce at least one million} \rverb!call! \rev{and} \rverb!return! \rev{events,
leading to overhead in excess of $10\times$. \name is smart enough to disable tracing if the
same program statement is encountered twice during a given execution, so that this statement
executes just as in vanilla Jupyter, while still detecting that symbol} \rverb!lst! \rev{should be
given} \rverb!f! \rev{as a dependency.}
\fi

\topic{\rev{Impact of Number of Cells on Analysis Latency}}
\rev{To better illustrate the benefit of using initialized analysis
for efficient computation of refresher cells, we measured the latency of just
\name's analysis component, and for each session, we plotted this time versus
the total number of cells created in the session, in \Cref{fig:analysis-vs-cells}.}

\frameme{\rev{{\bf \em Summary.}
While quadratic refresher computation is acceptable for sessions with relatively
few cells, we observe unacceptable per-cell latencies for larger notebooks with more than $50$
or so cells. The linear variant that leverages initialized analysis, however, scales gracefully
even for the largest notebooks in our execution logs.}}

\rev{The variance in \Cref{fig:analysis-vs-cells} for notebooks of the same size can be
attributed to cells with different amounts of code, as well as different numbers of cell
executions (since the size of the notebook is
\ifvldb
merely
\fi
a lower bound for the aforementioned
according to our replay strategy).}

\iftech
\subsection{\rev{Comparison with Other Systems}}
\label{sec:comparison:nb}

\begin{table}[t]
\begin{densecenter}
\rev{\thesissub{\scriptsize}{\small}\begin{tabular}{|c|c|c|c|c|}
\hline\rowcolor[HTML]{C5C5C5}{\bf System} & Datalore & Nodebook & Dataflow & \name \\\hline
Auto infers symbol lineage & \cmark & \cmark & \rxmark & \cmark \\\hline
Composes with \name       & \rxmark & \rxmark & \cmark & \na \\\hline
Auto resolves staleness & \cmark & \cmark & \cmark$^*$ & optionally \\\hline
\backref always does so correctly & \cmark & \rxmark & \cmark$^*$ & \na$^\ddag$ \\\hline
Preserves Jupyter semantics & \rxmark & \rxmark & \cmark$^\dag$ & \cmark \\\hline
No crashing on valid Python & \rxmark & \rxmark & \cmark & \cmark \\\hline
No other performance penalty    & \rxmark & \rxmark & \cmark & \rxmark{} (minor) \\\hline
\end{tabular}}
\caption{\rev{Summary of key distinguishing properties of notebook systems that help prevent stale executions. \\
{\scriptsize $*$ Only for manually specified dependencies. \\
$\dag$ Except for manually specified dependencies. \\
$\ddag$ \name can display the run-plan and allow manual corrections if needed.}}}
\label{tab:features}
\end{densecenter}
\end{table}

\rev{We now give a qualitative comparison of \name with other systems
that attempt to resolve staleness issues, viewed through the lens
of the data we collected in our empirical study.
After surveying relevant literature and open source software repositories,
we are aware of three such systems:
Dataflow notebooks~\cite{koop2017dataflow}, Nodebook~\cite{nodebook}, and the
Datalore kernel from JetBrains~\cite{datalore}.}

\rev{The most salient distinctions for each approach are summarized in \Cref{tab:features}.
We now provide a summary for each system.}

\emtitle{\rev{Dataflow Notebooks.}} \rev{While Dataflow notebooks have many desirable properties,
they require the user to specify dependencies manually in order to leverage any
staleness-resolving functionality. Dataflow notebooks can be used in conjunction with
\name if reactive cell execution via manually-specified dependencies is desired.}

\emtitle{\rev{Datalore and Nodebook.}} \rev{Both the Datalore kernel and Nodebook seem to take
take a hybrid analysis / memoization approach toward automatic staleness resolution:
each cell serializes the variables that it assigns, and if a cell $c$ is rerun, a liveness
checker determines what symbols need to be deserialized (using versions computed
by cells that appear in $c$ spatially) and used as ``inputs'' to $c$, possibly
rerunning cells prior to $c$ if they were edited or if they depend on an edited cell.
For example, in \Cref{fig:nblife:nb}, the second cell would be automatically rerun
if the user attempts to rerun the third cell after rerunning the first.}

\rev{These approaches allow notebooks to emulate script-like top-to-bottom behavior,
but serialization can come at significant cost for objects like large dataframes,
and furthermore, not all objects are serializable, thereby rendering these approaches
viable only on a much smaller set of programs, as we will see.
Finally, because liveness gives a conservative overestimation of symbols used,
these approaches may perform more work than necessary to rerun prior edited cells,
or to deserialize possibly-needed symbols.}

\topic{\rev{Ability to run valid Python}}
\rev{Perhaps the most serious shortcoming of memoization-based approaches
stems from their failure to execute valid Python code. Consider the
following\thesisdel{ example}:}
\begin{figure}[H]
\begin{lstjupyter}[cellnum={[1]}]{crash1}
  y = (i + 2 for i in range(10))
\end{lstjupyter}

\begin{lstjupyter}[cellnum={[3]}]{crash2}
  print(list(y))
\end{lstjupyter}
\end{figure}
\noindent \rev{If the user edits the first cell and then attempts to run the second
cell twice using either Nodebook or the Datalore kernel, they will
observe an error when these approaches try (and fail) to load the
non-serializable object} \rverb!y! \rev{from storage.}

\topic{\rev{Ability to conduct multiverse analyses}}
\rev{To facilitate the below discussion,
we define the ``rerun all cells'' approach adopted
by Nodebook, Datalore, and Dataflow notebooks (for manual dependencies)
to be a ``forcible cascade'' approach,
the selective rerun approach adopted by \name to be a
``supervised permissive cascade'' approach,
and that of Jupyter to be a ``manual cascade'' approach.}

\rev{In exploratory programming and data analysis, users do not usually
have a clear indication of which approach might work well up-front~\cite{kery2018interactions}.
So, they typically try various alternative approaches
to achieve their end-goal, while
also recording snippets of what they had tried previously,
for reuse, and for returning to old alternatives~\cite{kery2017variolite}.
The forcible cascade approach in this case
has the unintended effect of having all of their downstream alternatives
being executed, when the user wanted to execute precisely one.}

These sorts of multiverse analyeses are hindered by the forcible cascade
approaches. In fact, we found several instances in our execution logs
wherein users explicitly saved off variables to be returned to later,
and where forcible cascades would have overwritten these variables' saved
\iftech
values. We now give a typical example, depicted below:

\begin{figure}[H]
\begin{lstjupyter}[cellnum={[1]}]{casc1}
  df = pd.read_csv('universe1.csv')
\end{lstjupyter}

\begin{lstjupyter}[cellnum={[2]}]{casc2}
  df = df.dropna()[['col1', 'col2', 'col3', ...]]
\end{lstjupyter}

\begin{lstjupyter}[cellnum={[3]}]{casc3}
  df = df.rename({...})
\end{lstjupyter}

\begin{lstjupyter}[cellnum={[4]}]{casc4}
  df = df.merge(
           df.grouby(...).sum().reset_index(), 
           on='col', how='left')
\end{lstjupyter}

\begin{lstjupyter}[cellnum={[...]}]{cascdots}
  # more cells ...
\end{lstjupyter}

\begin{lstjupyter}[cellnum={[n]}]{cascn}
  df_saved = df.copy()
\end{lstjupyter}
\end{figure}

\rev{After reading the file} \rverb!universe1.csv! \rev{into a dataframe
and performing some transformations, the user would then save a
copy of the transformed dataframe. The user would then repeat
the same transformations by changing cell 1 to read in}
\rverb!universe2.csv! \rev{and manually running the cells below,
but stopping before creating the copy. The user then would
perform some comparison between the transformed} \rverb!universe1.csv!
\rev{data and the transformed} \rverb!universe2.csv! \rev{data.
Note that a forcible cascade would have overwritten the variable}
\rverb!df_saved!, \rev{preventing this comparison.}
\else
values; please see the technical report~\cite{nbtech} for more details.
\fi

\subsection{\rev{Staleness} Case Study}
\label{sec:case:nb}
\iftech
We now discuss a particularly egregious example of unsafe
behavior in one of the \numworking replayed sessions that
would have been caught by \name. In this session, the user
was attempting to visualize a Wiener process defined by
the following function:
\begin{figure}[H]
\begin{lstjupyter}[cellnum={[1]}]{wienerdef}
 def wiener(tmax, n):
    # Return one realization of a Wiener process
    # with n steps and a max time of tmax.
    times = np.linspace(0, tmax, n)    
    difference = np.diff(times)
    process = np.random.normal(0, difference**.5)
    process = np.cumsum(process)
    return times, process
\end{lstjupyter}
\end{figure}

\noindent The user then initially called this function and saved
the output in variables \verb!t! and \verb!w!:
\begin{figure}[H]
\begin{lstjupyter}[cellnum={[2]}]{wienercall1}
 t, w = wiener(1.0, 1000)
\end{lstjupyter}
\end{figure}

\noindent After inspecting a few values in the array \verb!w!, the user
then decided to rename it from lowercase \verb!w! to uppercase \verb!W!:
\begin{figure}[H]
\begin{lstjupyter}[cellnum={[3]}]{wienercall2}
 t, W = wiener(1.0, 1000)
\end{lstjupyter}
\end{figure}

Next, the user used the popular visualization library Altair~\cite{vanderplas2018altair}
to plot the output of the \verb!wiener! function, using the following code
(and producing output similar to the figure below the cell):
\begin{figure}[H]
\begin{lstjupyter}[cellnum={[4]}]{wienerplot1}
 data = pd.DataFrame({'time': t, !\HB!'W': w!\HE!})
 alt.Chart(data).mark_line().encode(
     x = 'time',
     y = 'W:Q'
 )
\end{lstjupyter}
\begin{tikzpicture}[remember picture, overlay]
\connectzm{wienerplot1-1}{orange}
\end{tikzpicture}
\begin{densecenter}
\resizebox{\linewidth}{!}{\input{\figs/wiener.pgf}}
\end{densecenter}
\end{figure}

\thesisdel{\vspace{-2em}}
\noindent However, note that in cell 4, the dataframe created
by the user, \verb!data!, refers
to the old lowercase \verb!w!, and not the
new uppercase \verb!W! most recently created. Thus,
when the user reran cells 3 and 4 in succession,
the exact same figure as that from the original cell 4 
was generated.

Confused, the user then reran cells 1, 3, and 4
in succession, each time generating a plot
identical to that from the original cell 4. This process
repeated itself around 20 times, before the user finally
noticed the problem and changed cell 4 to the following:
\begin{figure}[h]
\begin{lstjupyter}[cellnum={[4]}]{wienerplot2}
 data = pd.DataFrame({'time': t, !\HB!'W': W!\HE!})
 alt.Chart(data).mark_line().encode(
     x = 'time',
     y = 'W:Q'
 )
\end{lstjupyter}
\begin{tikzpicture}[remember picture, overlay]
\connectzm{wienerplot2-1}{orange}
\end{tikzpicture}
\end{figure}
\ifthesis
\vspace{1em}
\fi

This at last generated a different plot from the output of the original cell 4,
but the entire process resulted in a large amount of wasted effort
that would have been saved had the user noticed the error earlier.

\topic{How \name Helps}
To see how \name would have helped, let us examine the highlights
that \name would have presented after the user reran 3 and 4, got
confused, and then reran cell 1. The state of the notebook would
have then appeared similar to the following:
\begin{figure}[H]
\begin{lstjupyter}[cellnum={[7]}]{wienerdef2}
 def wiener(tmax, n):
    # Return one realization of a Wiener process
    # with n steps and a max time of tmax.
    times = np.linspace(0, tmax, n)    
    difference = np.diff(times)
    process = np.random.normal(0, difference**.5)
    process = np.cumsum(process)
    return (times, np.insert(process, 0, 0))
\end{lstjupyter}

\begin{lstjupyter}[cellnum={[5]},highlight=refresh]{wienercall3}
 t, W = wiener(1.0, 1000)
\end{lstjupyter}

\begin{lstjupyter}[cellnum={[6]},highlight=unsafe]{wienerplot4}
 data = pd.DataFrame({'time': t, 'W': w})
 alt.Chart(data).mark_line().encode(
     x = 'time',
     y = 'W:Q'
 )
\end{lstjupyter}
\end{figure}
\ifthesis
\vspace{1em}
\fi

\noindent That is, cell 3 would be given a \refresherhl{fresh}
cleanup suggestion highlight, because the \verb!wiener! symbol was recently updated when
the user reran the first cell (now labeled as 7). Likewise, cell 6
is given a \unsafehl{stale} highlight because lowercase
\verb!w! depends on the old version of \verb!wiener!.

Next, when the user reruns the cell labeled as 5 in the above
notebook, they would expect the unsafe highlight over cell 6
to be replaced with a fresh highlight, because they refreshed the symbol \verb!W!.
However, this does not occur, and the unsafe highlight remains.
The user could have then query \name's
API to determine why, and would have been presented with the following:
\begin{figure}[h]
\begin{lstjupyter}[cellnum={[7]}]{wienercall4}
 t, W = wiener(1.0, 1000)
\end{lstjupyter}

\begin{lstjupyter}[cellnum={[6]},highlight=unsafe]{unsafequery}
 data = pd.DataFrame({'time': t, 'W': w})
 alt.Chart(data).mark_line().encode(
     x = 'time',
     y = 'W:Q'
 )
 !\HB!
 # WARNING: `w` (latest update in cell 2) may depend
 #          on old version of symbol(s) [`wiener`].
                                                     !\HE!
\end{lstjupyter}
\begin{tikzpicture}[remember picture, overlay]
\connectzm{unsafequery-1}{red}
\end{tikzpicture}
\end{figure}
\ifthesis
\vspace{1em}
\fi

At this point, the user likely would have noticed that cell 6 does not refer to
symbol \verb!W!, which was updated when the user ran cell 7,
but on symbol \verb!w!, which
lingers in notebook state from when the user originally ran cell 2.
\else

In addition to \rev{the previous analyses}, we also sampled sessions
with safety errors in order to qualitatively study the severity of such errors.
\iftech
\else
We refer the reader to \rev{the} technical report~\cite{nbtech} \rev{for details}.
\fi
\fi

\fi

\section{Related Work}
\label{sec:related:nb}

Our work has connections to notebook systems, fine-grained
data versioning and provenance, and data-centric applications
of program analysis. \rev{Our notion of staleness
and cell execution orders is reminiscent
of the notion of serializability\techreport{. }\papertext{---we elaborate on this connection
in our technical report~\cite{nbtech}.}}
\techreport{ We survey each area below.}

\topic{Notebook Systems}
Error-prone interactions with global notebook state are well-documented
\techreport{in industry and academic
communities}~\cite{chattopadhyay2020s,idontlikenotebooks,head2019managing,kery2018story,koop2017dataflow,LauVLHCC2020,perkel2018jupyter,pimentel2019large,rule2018exploration,nodebook}.
The idea of treating a notebook as a dataflow computation graph
\techreport{with interdependent cells }has been studied previously~\cite{brachmann2020your,koop2017dataflow,nodebook};
however, \name is the first such system to our knowledge that preserves
existing any-order execution semantics.
We already surveyed Dataflow notebooks~\cite{koop2017dataflow}\rev{,}
Nodebook~\cite{nodebook}\rev{, and Datalore's kernel} in
\iftech
\Cref{sec:intro:nb,sec:comparison:nb}.
\else
\Cref{sec:intro:nb}.
\fi
\textsc{nbgather}~\cite{head2019managing} takes a purely static
\techreport{ and non-dataflow} approach
to automatically organize notebooks \techreport{using program
slicing techniques, and} thereby reducing non-reproducibility\techreport{ and errors due to messy notebooks}.
However, \techreport{the functionality provided by }\textsc{nbgather} \techreport{is orthogonal to \name and could be used in conjunction; for example,
it }does not help prevent state-related 
errors made before
\ifvldb
\textsc{nbgather}-induced
\fi
reorganization.

\topic{Versioning and Provenance}
\techreport{The work on data versioning and provenance has enjoyed a long history in the database community.}
Provenance capture
can be either {\em coarse-grained},
typically employed by scientific workflow systems, \eg~\cite{anand2009,bowers2012scientific,davidson2007provenance,davidson2008provenance,buneman2004archiving},
or {\em fine-grained} provenance as in database systems~\cite{green2007,cheney2009provenance,herschel2017survey},
typically at the level of individual rows.
\techreport{Within systems targeted toward individual scientists, Burrito~\cite{guo2012burrito} tracks file and script-level coarse-grained provenance, whereas our work falls under the fine-grained umbrella.}
\iftech
Within systems targeting scientific workflows, 
Burrito~\cite{guo2012burrito} tracks file and script-level coarse-grained provenance,
and noWorkflow~\cite{murta2014noworkflow,pimentel2017noworkflow} additionally captures finer-grained control flow dependencies, libraries,
and environment variables in scripts
as well as in computational notebooks~\cite{pimentel2015collecting}.
These systems target post-hoc analysis of fixed parameterized scripts to understand, \eg, how changing some
parameter affects the result of some experiment, but do not directly enable safer notebook interactions.
In contrast, the key observation of \nbs is that, in a notebook environment, written code is not
fixed up-front and often depends on the results of previously written code in a human-in-the-loop fashion;
we therefore leverage provenance to make it easier to reason about hidden notebook state in an online fashion,
providing hints and warnings to users as they go about their exploratory workflows.

Additional recent work
\else
Recent work
\fi
has examined challenges related to version compaction~\cite{bhattacherjee2015principles,huang2017orpheusdb}
and fine-grained lineage for scalable 
interactive visualization~\cite{psallidas2018smoke}\techreport{; our focus is
on enabling safer notebook interactions}.
\techreport{Toward this same end, }Vizier~\cite{brachmann2020your} attempts to
combine cell versioning and data provenance into a cohesive
notebook system with an intuitive interface, while warning
users of {\em caveats}
(\ie, possibly brittle assumptions that the analyst made about the data).
Like Vizier,
we leverage lineage to propagate information about potential errors.
However, data dependencies still need to be specified using
Vizier's dataset API, while \name infers them automatically\techreport{ using its tracer}.
\techreport{Furthermore, in our case, the semantics of the error stem directly from the ability to
execute cells out-of-order, while in Vizier, they stem from their so-called caveats.
That said, \name could, in principle, also propagate caveats;
incorporating an API for specifying such caveats is an interesting avenue for future work.}

\topic{Data-centric Program Checking}
The database community has traditionally leveraged program analysis to
{\em optimize database-backed applications}~\cite{emani2017dbridge,gupta2020aggify,ramachandra2017froid,yan2017understanding},
while we focus on catching bugs in an interactive notebook environment.
One exception is SQLCheck~\cite{dintyala2020sqlcheck}, which employs
a data-aware static analyzer to detect and fix so-called antipatterns that occur
during schema and query design. 
\techreport{Our goal with \name is similar in spirit,
though we focus on detecting and rectifying potential errors that occur over the
course of interactive notebook sessions.
Within the notebook space,
Vizier~\cite{brachmann2020your} also uses static analysis to determine whether particular queries are affected by brittle assumptions / caveats.
This use case is orthogonal to our goal, which is to preserve traditional notebook semantics
while reducing error-proneness of such interactions;
we could incorporate caveat-checking into \name's static analysis in the future
(by, \eg, detecting liveness of symbols with attached caveats in cells).}
\smacke{Make sure opt refs are representative / comprehensive}

\iftech
\topic{\rev{Parallels with Transaction Processing}}
\rev{There are some parallels between the notion of transactional
serializability and safety.
For example, if we view a cell as a transaction,
conflicts between two transactions would correspond
to dependencies between cells (either in a R/W, W/R or W/W fashion).
Moreover, our goal in identifying
stale and refresher cells is akin to {\em conservatively} identifying
whether the execution order corresponds to a safe / desirable schedule in
terms of the read/writes (\eg, whether the schedule is view equivalent to a
``run from top-to-bottom schedule'').
Such conservative mechanisms of identifying schedules that adhere
to various consistency have been proposed in prior work~\cite{roy2015homeostasis,zhang2013transaction}.
However, the similarities largely end there:}
\begin{denseenum}
\item \rev{If we view a cell as a transactional boundary,
reads and writes within one cell cannot be interleaved
with reads and writes within another cell.
Thus, the notion of serializability is itself too weak in that
it allows for interleavings between transactions.}
\item \rev{Say we abandon the notion of serializability,
but instead consider the notion of view equivalence
of two different cell execution schedules.
Here, we note that there are often multiple ways
to "refresh" a stale cell, typical in multiverse 
analyses---corresponding to different
execution paths in a DAG of cell dependencies (or lineage).
Any one of these would be permissible from our viewpoint.
On the other hand, view equivalence is a strict 
linear definition, unlike our DAG-based permissive definition.}
\item \rev{Finally, even in the case that there is a single
path in our DAG, view equivalence ends up being overly conservative,
dismissing certain valid cell execution schedules 
as non-equivalent, when they are indeed equivalent 
from an end result standpoint.}
\end{denseenum} 
\rev{To illustrate this last point, consider the following three cells:}

\begin{figure}[H]
\begin{lstjupyter}[cellnum={[1]}]{reveq1}
  x = 0
\end{lstjupyter}

\begin{lstjupyter}[cellnum={[2]}]{reveq2}
  y = 5
  print(x)
\end{lstjupyter}

\begin{lstjupyter}[cellnum={[3]}]{reveq3}
  print(y)
\end{lstjupyter}
\end{figure}

\rev{In the above example, $c_1$ writes $x$, $c_2$ reads $x$ and writes $y$,
and cell $c_3$ reads $y$. Suppose the user changes the assignment in $c_1$
to} \rverb!x = 42!. \rev{If enforcing view serializability, we would highlight cell $c_3$
as unsafe to execute, because $c_2$ would need to execute before it under
view serializability: $c_2$ reads $c_1$'s write of $x$, and $c_3$ in turn reads $c_2$'s
write of $y$. However, it is easy to see in the above example that the order of re-execution
between $c_2$ and $c_3$ does not matter. If, on the other hand, $c_2$ had set}
\rverb!y = x + 1!, \rev{then it is clear that $c_2$ should be rerun before $c_3$ can be
safely rerun. It is for this reason that we adopted a lineage-centric framework,
wherein cells are used primarily to associate timestamps with symbols
and blocks of code via their execution counters.}
\fi

\section{Conclusion}
\label{sec:conclusion:nb}

We presented \name, a kernel and frontend for Jupyter that
attempts to detect and correct potentially unsafe
interactions in notebooks, all while preserving the flexibility
of familiar any-order notebook semantics. We described the implementation of
\name's tracer, checker, and frontend, and how they integrate into existing
notebook workflows to {\em efficiently} reduce error-proneness in notebooks.
We showed how cells that \name would have warned as unsafe
were actively avoided, and cells that would have been suggested for re-execution were prioritized
by real users on a corpus of \numworking real notebook sessions.
\techreport{While we focused on unsafe interactions due to staleness in this \work,
extending our approach to other types of unsafe interactions is a promising
direction for future research.}
\iftech
\else

\begin{acks}
{\footnotesize We thank the anonymous reviewers for their valuable feedback.
We additionally thank Sarah Chasins for bringing to our attention the connection between refresher cells and
initialized variable analysis.
We acknowledge support from grants IIS-1940759 and IIS-1940757 awarded by the National Science Foundation, and funds from the Alfred P. Sloan Foundation, Facebook, Adobe, Toyota Research Institute, Google, the Siebel Energy Institute, and State Farm.
The content is solely the responsibility of the authors and does not necessarily represent the official views of the funding agencies and organizations.}
\end{acks}
\fi


\ifvldb
\else
\bibliographystyle{ACM-Reference-Format}
\balance
\bibliography{main,nb,provrefs,analysis}
\fi



\ifvldb
\balance
\printbibliography[title=REFERENCES]
\else
\makeatletter
\let\ACM@origbaselinestretch\baselinestretch
\makeatother
\fi
\end{document}